\documentclass[a4paper]{article}
\usepackage{amsmath}
\usepackage{amssymb}
\usepackage[a4paper,includeheadfoot]{geometry}

\input{tcilatex}
\begin{document}

\title{The observable-state model and non-renormalizable theories}
\author{Juan Sebasti\'{a}n Ardenghi\thanks{%
email:\ jsardenghi@gmail.com, fax number:\ +54-291-4595142} \\
IFISUR, Departamento de F\'{\i}sica (UNS-CONICET)\\
Avenida Alem 1253, Bah\'{\i}a Blanca, Buenos Aires, Argentina \and Alfredo
Juan \\
IFISUR, Departamento de F\'{\i}sica (UNS-CONICET)\\
Avenida Alem 1253, Bah\'{\i}a Blanca, Buenos Aires, Argentina \and Mario
Castagnino \\
Institutos de F\'{\i}sica de Rosario y de Astronom\'{\i}a y F\'{\i}sica del
Espacio.\\
Casilla de Correos 67, sucursal 28, 1428 Buenos Aires, Argentina.}
\maketitle

\begin{abstract}
The aim of this work is to apply the observable-state model for the quantum
field theory of a $\phi ^{n}$ self-interaction. We show how to obtain finite
values for the $2$-point and $n$-point correlation functions without
introducing counterterms in the Lagrangian. Also, we show how to obtain the
renormalization group equation for the mass and the coupling constant.
Finally, we found the dependence of the coupling constant with the energy
scale and we discuss the validity of the observable-state model in terms of
the projection procedure.
\end{abstract}

\section{Introduction}

This paper is the application of the projection method called "The
observable-state model" introduced in papers \cite{PR1} and \cite{PR2}, to
the non-renormalizable $\phi ^{l}$ theories, with $l>4.$

The non-renormalizable theories have not been extensively studied because it
is usual to believe that they do not contain physical predictions.
Nevertheless, they can be studied as effective theories, where the
predictive power decay at energies of the order of the common mass $M$ that
characterizes the various couplings (see \cite{W}, page 523). In turn, the
effective quantum field theories are very useful because the short distance
features can be ignored producing an enormous simplification, where only the
light particles are important (see \cite{Georgi}). But in the projection
method it is not necessary to introduce counterterms in the Lagrangian to
cancel the divergences due to the short-distance interactions that appear in
the perturbation expansion. Therefore, the renormalizable and
non-renormalizable theories are on equal footing from the point of view of
the projection procedure. In \cite{PR2} we have already shown that the
finite contribution to the two and fourth correlation functions in $\phi
^{4} $ theory gives the correct renormalization group equations at one-loop
correction. So it makes sense to apply the projection procedure to the $\phi
^{l}$ theory to obtain the corresponding mass and coupling constant
renormalization group equations at one loop. Basically, the projection
method work as follows:\ we write the generating functional $Z[J]$ for the
correlation functions of $n$ external points as a mean value of an
observable, defined through the $J(x_{i})$ external sources, in a quantum
state defined by the correlation function itself. Then, by making a
equivalence class on the quantum states, we can separate the short-distance
behavior of the physics which appears in the diagonal part of the quantum
state written in the coordinate basis. In this way, the divergences show up
as Dirac deltas that are represented in a dimensional regularization scheme
by the poles $\frac{1}{\epsilon ^{k}}$, where $\epsilon =d-4$ and $d$ is the
space-time dimension. Then we can simply disregard these unphysical
infinities by applying a projection operator on the quantum states that
remove the diagonal part. A conceptual discussion of the observable-state
model has been introduced in Section VI of \cite{PR2}.\footnote{%
In the next section a short summary of the observable-state model is
introduced.}

The results found in this work not only can be applied to the
non-renormalizable theories in Quantum Field Theory, but also to condensed
matter system, where the thermal fluctuations are correlated only over
atomic distances and under special circunstances, over longer ranges. The
behavior of any statistical system under a second-order phase transition can
be translated into the behavior of a particular quantum field theory. A
concrete example is exhibited by a ferromagnet \cite{landau}. In this case,
the quantum field represents the local spin density $s(x)$ and the
self-interaction of this quantum field can be obtained by expanding the
Gibbs free energy in terms of $s(x)$. Then, the successive approximations
are introduced as even powers of $s(x)$. In general, terms of order $s^{6}$
or higher are ignored because $s$ is small. But in the general case, the
contributions of these higher orders must be computed. In this sense, this
work contributes to this calculation under the renormalization method
introduced in the previous works \cite{PR1} and \cite{PR2}.

The paper is organized as follows:

In section II a brief review of the observable-state model is introduced.

In section III we compute the first correction to the two-point correlation
function in a\textit{\ }$\phi ^{l}$ theory with the subsequent mass
renormalization group equation. We show an example with $l=6$, which is a
non-renormalizable theory in a space-time with dimension $d=4$.

In section IV we compute the second correction to the $l$-correlation
function in a $\phi ^{l}$ theory. We compute the coupling constant
renormalization group equation and we show an example with $l=6$.

In section V we present a discussion about the mass renormalization group
and the renormalization conditions. Finally, in section VI we present the
conclusions and in the Appendix A we show a detailed computation of the
second order in the perturbation expansion of the $l$-correlation function
of a $\phi ^{l}$ theory.

\section{Observables and states in quantum field theory: the main idea}

The starting point of the observable-state model is some (symmetric) $n$%
-point functions $\tau ^{(n)}(x_{1},...,x_{n})$ (like Feynman or Euclidean
functions), and its corresponding generating functional (\cite{Haag}, eq.
(II.2.21), \cite{Brown}, eq. (3.2.11)). Then, the main equation reads:

\begin{equation}
iZ\left[ J\right] =\underset{n=0}{\overset{\infty }{\sum }}\underset{p=0}{%
\overset{\infty }{\sum }}\frac{i^{n}}{n!}\frac{i^{p}}{p!}\int \left\langle
\Omega _{0}\left\vert T\phi _{0}(x_{1})...\phi _{0}(x_{n})\mathcal{L}%
_{I}^{0}(y_{1})...\mathcal{L}_{I}^{0}(y_{p})\right\vert \Omega
_{0}\right\rangle J(x_{1})...J(x_{n})\underset{i=1}{\overset{n}{\prod }}%
d^{4}x_{i}\underset{i=1}{\overset{p}{\prod }}d^{4}y_{i}  \label{ideas7}
\end{equation}%
where $y_{i}$ are the internal vertices of the perturbation expansion and $%
\mathcal{L}_{I}^{0}(y_{p})$ is the Lagrangian interaction density (see
eq.(II.2.33) of \cite{Haag}).

The generating functional $Z[J]$ can be written as an mean value of an
observable defined through the $J(x_{n})$ sources in a quantum state defined
by the correlation function $\left\langle \Omega _{0}\left\vert T\phi
(x_{1})...\phi (x_{n})\mathcal{L}_{I}^{0}(y_{1})...\mathcal{L}%
_{I}^{0}(y_{p})\right\vert \Omega _{0}\right\rangle $.\footnote{%
In some sense, these observables will be the particle detector (see \cite%
{Thnew}, page 6, below eq.(2.6)).} This procedure will be done for each
correlation function of $n$ external points.

Using dimensional regularization (see \cite{TW}) we can write the
one-particle irreducible contribution to the correlation function such that
(see \cite{critical} for $\phi ^{4}$ theory):

\begin{equation}
\int \left\langle \Omega _{0}\left\vert T\phi (x_{1})...\phi (x_{n})\mathcal{%
L}_{I}^{0}(y_{1})...\mathcal{L}_{I}^{0}(y_{p})\right\vert \Omega
_{0}\right\rangle \underset{i=1}{\overset{p}{\prod }}%
d^{4}y_{i}=f_{0}^{(n)}(x_{1},...,x_{n})\overset{+\infty }{\underset{l=-L(n,p)%
}{\sum }}\beta _{l}^{(n,p)}(m_{0}^{2},\mu )\epsilon ^{l}  \label{la1}
\end{equation}%
where $f_{0}^{(n)}$ is some function of the external points, $\beta
_{l}^{(n,p)}(m_{0}^{2},\mu )$ are some coefficients of the dimensional
regularization that depends on the external momentum, the mass factor $\mu $
used to keep the coupling constant dimensionless and the mass of the field $%
m_{0}$. The parameter $\epsilon $ is $\epsilon =d-4$, where $d$ is the
dimension of space-time. The sum in $l$ starts at $-L(n,p)$ where $L(n,p)$
is the number of loops at order $p$ in the correlation functions of $n$
external points (see Appendix A, eq.(A6) of \cite{PR1}). The functions $%
f_{0}^{(n)}$ and $L(n,p)$ for $\phi ^{4}$ theory reads%
\begin{equation}
f_{0}^{(n)}=\overset{n}{\underset{i=1}{\prod }}\int \frac{d^{4}p_{i}}{(2\pi
)^{4}}\frac{e^{-ip_{i}x_{i}}}{p_{i}^{2}-m_{0}^{2}}\delta (\underset{j=1}{%
\overset{n}{\sum }}p_{j}^{2})\text{, \ \ \ \ \ \ \ \ \ \ \ \ \ \ \ \ \ \ \ \ 
}L(n,p)=p-\frac{n}{2}+1  \label{la5}
\end{equation}%
Inserting eq.(\ref{la1}) in eq.(\ref{ideas7}) we obtain\footnote{%
The infinite sum in the $l$ index in eq.(\ref{la1}) can be truncated in $l=0$%
, because the remaining terms are proportional to $\epsilon ^{l}$ and the
final result must be computed by taking the $\epsilon \rightarrow 0$ limit.
In this sense, what concern us is the principal part plus the constant term
of the Laurent serie with poles $d-4$.}%
\begin{equation}
iZ\left[ J\right] =\underset{n=0}{\overset{\infty }{\sum }}\underset{p=0}{%
\overset{\infty }{\sum }}\frac{i^{n}}{n!}\frac{i^{p}}{p!}\overset{0}{%
\underset{l=-L(n,p)}{\sum }}\beta _{l}^{(n,p)}\epsilon ^{l}\int
f_{0}^{(n)}(x_{1},...,x_{n})J(x_{1})...J(x_{n})\underset{i=1}{\overset{n}{%
\prod }}d^{4}x_{i}  \label{la6}
\end{equation}%
The observable-state model consist in the assumption that the generating
functional of last equation can be rewritten as a mean value of the
following observable

\begin{equation}
O^{(n,p)}=O_{ext}^{(n)}\otimes I_{int}^{(p)}  \label{lala0}
\end{equation}%
in the following quantum state%
\begin{equation}
\rho ^{(n,p)}=\rho _{ext}^{(n)}\otimes \rho _{int}^{(n,p)}  \label{lala0.1}
\end{equation}%
where $O_{ext}^{(n)}$ is some observable that acts on the external
coordinates $x_{i}$ and $I_{int}^{(p)}$ is the identity operator that acts
on the internal vertices due to the perturbation expansion. In a similar
way, $\rho _{ext}^{(n)}$ is the quantum state of the external part and $\rho
_{int}^{(n,p)}$ is the quantum state of the internal part.

Then, the mean value of $O^{(n,p)}$ in $\rho ^{(n,p)}$ reads%
\begin{equation}
Tr(\rho ^{(n,p)}O^{(n,p)})=Tr(\rho _{ext}^{(n)}O_{ext}^{(n)})Tr(\rho
_{int}^{(n,p)})  \label{lala0.2}
\end{equation}%
Using last equation, the generating functional of eq.(\ref{la6}) can be
written as%
\begin{equation}
iZ\left[ J\right] =\underset{n=0}{\overset{\infty }{\sum }}\underset{p=0}{%
\overset{\infty }{\sum }}\frac{i^{n}}{n!}\frac{i^{p}}{p!}Tr(\rho
^{(n,p)}O^{(n,p)})=\underset{n=0}{\overset{\infty }{\sum }}\underset{p=0}{%
\overset{\infty }{\sum }}\frac{i^{n}}{n!}\frac{i^{p}}{p!}Tr(\rho
_{int}^{(n,p)})Tr(\rho _{ext}^{(n)}O_{ext}^{(n)})  \label{lala0.3}
\end{equation}%
where%
\begin{equation}
\rho _{ext}^{(n)}=\int f_{0}^{(n)}(x_{1},...,x_{n})\left\vert x_{1},...,x_{%
\frac{n}{2}}\right\rangle \left\langle x_{\frac{n}{2}+1},...,x_{n}\right%
\vert \underset{i=1}{\overset{n}{\prod }}d^{4}x_{i}  \label{lala0.4}
\end{equation}%
and%
\begin{equation}
O_{ext}^{(n)}=\int J(x_{1})...J(x_{n})\left\vert x_{1},...,x_{\frac{n}{2}%
}\right\rangle \left\langle x_{\frac{n}{2}+1},...,x_{n}\right\vert \underset{%
i=1}{\overset{n}{\prod }}d^{4}x_{i}  \label{lala0.5}
\end{equation}%
In turn%
\begin{equation}
Tr(\rho _{int}^{(n,p)})=\overset{+\infty }{\underset{l=-L(n,p)}{\sum }}\beta
_{l}^{(n,p)}\epsilon ^{l}  \label{lala0.6}
\end{equation}%
which implies that the divergences of the quantum field theory are the
consequence of taking the trace of the internal quantum state $\rho
_{int}^{(n,p)}$. This point is relevant; because the trace of an operator is
an invariant quantity, this means that it is the same in different basis.
This implies that if we want to obtain a finite contribution $\beta
_{0}^{(n,p)}$, we must apply a non-unitary transformation on $\rho
_{int}^{(n,p)}$ that changes its trace, i.e., we must project to another $%
\rho _{int}$.

\subsection{Internal quantum state}

To define the internal quantum state we will just recall some considerations
(see Section VI in \cite{PR1}): the algebra of observables $\mathcal{O}$ is
represented by $^{\ast }-$algebra $\mathcal{A}$ of self-adjoint elements and
states are represented by functionals on $\mathcal{O}$, that is, by elements
of the dual space $\mathcal{O}^{\prime }$, $\rho \in \mathcal{O}^{\prime }$.
We will construct a $C^{\ast }-$algebra of operators defined in terms of
elements with the property $Tr(A^{\ast }A)<\infty $. As it is well known, a $%
C^{\ast }-$algebra can be represented in a Hilbert space $\mathcal{H}$ (GNS\
theorem)\footnote{%
Gelfand, Naimark and Segal \cite{GNS}.} and, in this particular case $%
\mathcal{O=\mathcal{O}}^{\prime }$; therefore$\mathcal{\ \mathcal{O}}$ and $%
\mathcal{O}^{\prime }$ are represented by $\mathcal{H\otimes \mathcal{H}}$
that will be called $\mathcal{N}$, the Liouville space.

As we are interested in the diagonal and non-diagonal elements of a matrix
state we can define a sub algebra of $\mathcal{\mathcal{\mathcal{N}}}$, that
can be called a van Hove algebra (\cite{van hove}, \cite{van hove1}, \cite%
{van hove2}, \cite{van hove3}, \cite{van hove4}) since such a structure
appears in his work as:%
\begin{equation}
\mathcal{\mathcal{\mathcal{N}}}_{vh}\mathcal{=N}_{S}\oplus \mathcal{N}%
_{R}\subset \mathcal{\mathcal{\mathcal{N}}}  \label{k}
\end{equation}%
where the vector space $\mathcal{N}_{R}$ is the space of operators with $%
O(x)=0$ and $O(x,x^{\prime })$ is a regular function. Moreover $\mathcal{O=N}%
_{vhS}$ is the space of selfadjoint operators of $\mathcal{\mathcal{\mathcal{%
N}}}_{vh},$ which can be constructed in such a way it could be dense in $%
\mathcal{N}_{S}$ (because any distribution can be approximated by regular
functions) (for the details see \cite{PR1}, Section II.B and Section VI).
Therefore essentially the introduced restriction is the minimal possible
coarse-graining. Now the $\oplus $ is a direct sum because $\mathcal{N}_{S}$
contains the factor $\delta (x-x^{\prime })$ and $\mathcal{N}_{R}$ contains
just regular functions and a kernel cannot be both a $\delta $ and a regular
function. Moreover, as our observables must be self-adjoint, the space of
observables must be 
\begin{equation}
\mathcal{O=N}_{vhS}\mathcal{=N}_{S}\oplus \mathcal{N}_{R}\subset \mathcal{N}
\label{l}
\end{equation}

The states must be considered as linear functionals over the space $\mathcal{%
O}$ ($\mathcal{O}^{\prime}$ the dual of space $\mathcal{O}$):

\begin{equation}
\mathcal{O}^{\prime }\mathcal{=N}_{vhS}^{\prime }\mathcal{=N}_{S}^{\prime
}\oplus \mathcal{N}_{R}^{\prime }\subset \mathcal{N}^{\prime }  \label{m}
\end{equation}%
The set of these generalized states is the convex set $\mathcal{S\subset O}%
^{\prime }$.

Having this in mind, we can define the internal quantum state in the
following way%
\begin{eqnarray}
\rho _{int}^{(n,p)} &=&\int \underset{i=1}{\overset{L(n,p)}{\prod }}\left(
\rho _{D}^{(n,p,i)}(y_{i})\delta (y_{i}-w_{i})+\rho
_{ND}^{(n,p,i)}(y_{i},w_{i})\right)  \label{c9.1} \\
&&\left\vert y_{1},...,y_{L(n,p)}\right\rangle \left\langle
w_{1},..,w_{L(n,p)}\right\vert \underset{i=1}{\overset{L(n,p)}{\prod }}%
d^{4}y_{i}d^{4}w_{i}  \notag
\end{eqnarray}%
\bigskip

The trace reads\footnote{%
In eq.(\ref{c9.3}) we have introduced an equivalence between the Dirac delta
and the pole parameter of the dimensional regularization that has been shown
in Appendix A of \cite{PR2}.}%
\begin{equation}
Tr(\rho _{int}^{(n,p)})=\underset{i=1}{\overset{L(n,p)}{\prod }}\left( \frac{%
\rho _{D}^{(n,p,i)}}{\pi \epsilon }+\rho _{ND}^{(n,p,i)}\right)  \label{c9.3}
\end{equation}%
where%
\begin{equation}
\rho _{D}^{(n,p,i)}=\int\limits_{{}}^{{}}\rho _{D}^{(n,p,i)}(y_{i})d^{4}y_{i}%
\text{ \ \ \ \ \ \ \ \ }\rho _{ND}^{(n,p,i)}=\int\limits_{{}}^{{}}\rho
_{ND}^{(n,p,i)}(y_{i},y_{i})d^{4}y_{i}  \label{c9.4}
\end{equation}%
We can see from last equation that $\rho _{D}^{(n,p,i)}$ and $\rho
_{ND}^{(n,p,i)}$ are merely normalization factors. Eq.(\ref{c9.3}) can be
written as%
\begin{equation}
Tr(\rho _{int}^{(n,p)})=\underset{l=-L(n,p)}{\overset{0}{\sum }}\gamma
_{l}^{(n,p)}\epsilon ^{l}  \label{c9.50}
\end{equation}%
where%
\begin{equation}
\gamma _{0}^{(n,p)}=\underset{i=1}{\overset{L(n,p)}{\prod }}\rho
_{ND}^{(n,p,i)}\text{, ... \ , \ \ }\gamma _{L(n,p)}^{(n,p)}=\frac{1}{\pi
^{L(n,p)}}\underset{i=1}{\overset{L(n,p)}{\prod }}\rho _{D}^{(n,p,i)}
\label{c9.501}
\end{equation}

All the terms $\gamma _{l}^{(n,p)}$ with $l>0$ that are multiplied by $%
\epsilon ^{l}$ contain at least one $\rho _{D}^{(n,p,i)}$, that is, the
diagonal part of the state of the $\dot{i}-$internal quantum system. In
particular, we can make the following equality%
\begin{equation}
\beta _{l}^{(n,p)}=\gamma _{l}^{(n,p)}  \label{c9.7}
\end{equation}

In this sense, the coefficients obtained by the dimensional regularization
can be associated with the products of the diagonal and non-diagonal parts
of the internal quantum state. In particular, the coefficient that is not
multiplied by a $\epsilon $ is $\gamma _{0}^{(n,p)}$ which depends
exclusively on the non-diagonal quantum state.

\subsection{Projection over the finite contribution}

As we saw in eq.(\ref{c9.50}) and eq.(\ref{c9.501}), the finite result
exclusively depends on the non-diagonal quantum state, so we can construct a
projector that projects over the non-diagonal quantum state. This projector
reads\footnote{%
Is not difficult to show that it is a projector: linearity implies that $\Pi
(a+b)=\Pi (a)+\Pi (b)$, then, if $\Pi (a)=a-G$, then, $\Pi ^{2}(a)=\Pi
(a-G)=\Pi (a)-\Pi (G)$, but $\Pi (G)=G-G=0$, then $\Pi ^{2}(a)=\Pi (a)$.}%
\begin{eqnarray}
\Pi _{p}(\rho _{int}^{(n,p)})=\rho _{int}^{(n,p)}-\int \rho
_{D}^{(n,p,1)}(y_{1})\rho _{D}^{(n,p,2)}(y_{2})...\rho
_{D}^{(n,p,L(n,p))}(y_{L(n,p)})\left\vert y_{1},...,y_{L(n,p)}\right\rangle
\left\langle y_{1},..,y_{L(n,p)}\right\vert \underset{i=1}{\overset{L(n,p)}{%
\prod }}d^{4}y_{i}  \label{finite1} \\
+\int \rho _{D}^{(n,p,1)}(y_{1})\rho _{D}^{(n,p,2)}(y_{2})...\rho
_{D}^{(n,p,L(n,p)-1)}(y_{L(n,p)-1})\rho
_{ND}^{(n,p,L(n,p))}(y_{L(n,p)},w_{L(n,p)})  \notag \\
\left\vert y_{1},...,y_{L(n,p)}\right\rangle \left\langle
y_{1},..,w_{L(n,p)}\right\vert d^{4}w_{L(n,p)}\underset{i=1}{\overset{%
L(n,p)-1}{\prod }}d^{4}y_{i}+...+\int \rho _{D}^{(n,p,1)}(y_{1})\rho
_{ND}^{(n,p,2)}(y_{2},w_{2})...\rho
_{ND}^{(n,p,L(n,p))}(y_{L(n,p)},w_{L(n,p)})  \notag \\
\left\vert y_{1},...,y_{L(n,p)}\right\rangle \left\langle
y_{1},..,w_{L(n,p)}\right\vert d^{4}y_{1}\underset{i=2}{\overset{L(n,p)}{%
\prod }}d^{4}y_{i}d^{4}w_{i})  \notag
\end{eqnarray}%
The projection procedure consists in the subtraction of the part of the\
state that contains at least one internal diagonal quantum state. This
projector acting on the state $\rho ^{(n,p)}$ yields%
\begin{equation}
\Pi _{p}(\rho _{int}^{(n,p)})=\int \underset{i=1}{\overset{L(n,p)}{\prod }}%
\rho _{ND}^{(n,p,i)}(y_{i},w_{i})\left\vert
y_{1},...,y_{L(n,p)}\right\rangle \left\langle
w_{1},..,w_{L(n,p)}\right\vert \underset{i=1}{\overset{L(n,p)}{\prod }}%
d^{4}y_{i}d^{4}w_{i}  \label{finite2}
\end{equation}%
Then, using the equivalence of eq.(\ref{c9.7}), the mean value of $O^{(n,p)}$
in the state $\Pi _{p}(\rho ^{(n,p)})$ reads:%
\begin{equation}
Tr(\Pi _{p}(\rho ^{(n,p)})O^{(n,p)})=\beta _{0}^{(n,p)}\int
f_{0}^{(n)}(x_{1},...,x_{n})O_{ext}^{(n)}\left( x_{1},...,x_{n}\right) 
\underset{i=1}{\overset{n}{\prod }}d^{4}x_{i}  \label{finite3}
\end{equation}%
where $O_{ext}^{(n)}\left( x_{1},...,x_{n}\right) =J(x_{1})...J(x_{n})$ (see
eq.(\ref{lala0.5})). Multiplying by $\frac{i^{p}}{p!}$ and summing in $p$ we
obtain\footnote{%
The factor $\frac{i^{p}}{p!}$ is introduced for later convenience, but its
meaning could be that in the observable-state model, the quantum state is
invariant under an exchange of internal vertices.}%
\begin{equation}
Tr(\rho ^{(n)}O_{ext}^{(n)})=\overset{+\infty }{\underset{p=0}{\sum }}\frac{%
i^{p}}{p!}Tr(\Pi _{p}(\rho ^{(n,p)})O^{(n,p)})=\overset{+\infty }{\underset{%
p=0}{\sum }}\frac{i^{p}}{p!}\beta _{0}^{(n,p)}\int
f_{0}^{(n)}(x_{1},...,x_{n})O_{ext}^{(n)}\left( x_{1},...,x_{n}\right) 
\underset{i=1}{\overset{n}{\prod }}d^{4}x_{i}  \label{finite3.1}
\end{equation}%
where%
\begin{equation}
\rho ^{(n)}=\left( \overset{+\infty }{\underset{p=0}{\sum }}\frac{i^{p}}{p!}%
\beta _{0}^{(n,p)}\right) \rho _{ext}^{(n)}  \label{finite3.2}
\end{equation}%
where $\frac{i^{p}}{p!}\beta _{0}^{(n,p)}$ is the coefficient of the quantum
state $\rho _{ext}^{(n)}$.

In this way, we can eliminate all the divergences of the observable-state
model by the application of the projector over a well defined Hilbert
subspace. This formalism has been applied to the correlation functions of
\thinspace $n=0$, $n=2$ and $n=4$ external points for $\phi ^{4}$ theory
(see \cite{PR1} and \cite{PR2}). The main idea of this work is to apply the
same procedure to the correlation function of $n=2~$and $n=l$ external
points of a $\phi ^{l}$ self-interaction.

\section{First correction to mass renormalization in $\protect\phi ^{l}$
theories}

To obtain the first correction to the mass renormalization with a $\phi ^{l}$
self interaction, we must expand in a perturbation expansion the two-point
correlation function:%
\begin{equation}
\left\langle \Omega \left\vert \phi (x_{1})\phi (x_{2})\right\vert \Omega
\right\rangle =\left\langle \Omega _{0}\left\vert \phi _{0}(x_{1})\phi
_{0}(x_{2})\right\vert \Omega _{0}\right\rangle +(-i\lambda _{0})\dint
\left\langle \Omega _{0}\left\vert \phi _{0}(x_{1})\phi _{0}(x_{2})\phi
_{0}^{l}(y_{1})\right\vert \Omega _{0}\right\rangle d^{4}y_{1}+...
\label{new1}
\end{equation}%
where $\left\vert \Omega \right\rangle $ and $\phi (x)$ are the vacuum and
the quantum field of the interacting theory, $\left\vert \Omega
_{0}\right\rangle $ and $\phi _{0}(x)$ are the vacuum and the quantum field
of the non-interacting theory. The correction at first order can be written
from the observable-state model viewpoint as%
\begin{equation}
G^{(2,l,1)}(x_{1},x_{2})=(-i\lambda _{0})\dint \left\langle \Omega
_{0}\left\vert \phi _{0}(x_{1})\phi _{0}(x_{2})\phi
_{0}^{l}(y_{1})\right\vert \Omega _{0}\right\rangle d^{4}y_{1}=\rho
_{ext}^{(2)}(x_{1},x_{2})Tr(\rho _{int}^{(l,2,1)})  \label{new2}
\end{equation}%
where%
\begin{equation}
\rho _{ext}^{(2)}(x_{1},x_{2})=\dint\limits_{{}}^{{}}\frac{d^{4}p}{(2\pi
)^{4}}\frac{e^{-ip(x_{1}-x_{2})}}{(p^{2}-m_{0}^{2})^{2}}  \label{new2.0}
\end{equation}%
and (see Section II of \cite{PR2})%
\begin{equation}
Tr(\rho _{int}^{(l,2,1)})=-\lambda _{0}^{(l)}\left[ \Delta (0)\right] ^{%
\frac{l}{2}-1}  \label{new.2.1}
\end{equation}%
where the superscript $l$,\thinspace\ $2$ and $1$ refers to the power of the
interaction, to the first order in the perturbation and to the two-point
correlation function respectively. The $\rho _{ext}^{(2)}(x_{1},x_{2})$
function is the coefficient of the following quantum state%
\begin{equation}
\rho _{ext}^{(2)}=\dint \rho _{ext}^{(2)}(x_{1},x_{2})\left\vert
x_{1}\right\rangle \left\langle x_{2}\right\vert d^{4}x_{1}d^{4}x_{2}
\label{new2.2}
\end{equation}%
and $\Delta (0)$ is the scalar propagator valuated in zero. This propagator
is singular, using dimensional regularization (see \cite{TW}) we have 
\begin{equation}
\Delta (0)=\underset{j=-1}{\overset{+\infty }{\sum }}\alpha _{j}\epsilon
^{j}=\frac{\alpha _{-1}}{\epsilon }+\alpha _{0}+\alpha _{1}\epsilon +...
\label{new3}
\end{equation}%
where $\epsilon =d-4$ and $d$ is the dimension of space-time. The low terms
reads%
\begin{gather}
\alpha _{-1}=-\frac{m_{0}^{2}}{8\pi ^{2}}  \label{new4} \\
\alpha _{0}=\frac{m_{0}^{2}}{16\pi ^{2}}\left( 1-\gamma +\ln (\frac{4\pi }{%
m_{0}^{2}})\right)  \notag \\
\alpha _{1}=\frac{m_{0}^{2}}{384\pi ^{2}}\left( 6\ln (\frac{m_{0}^{2}}{4\pi }%
)\left( 2\gamma -2+\ln (\frac{m_{0}^{2}}{4\pi })\right) +\pi ^{2}+6\gamma
^{2}-12\gamma +12\right)  \notag
\end{gather}%
and all the $\alpha _{j}$ functions depends on $m_{0}^{2}$.

Now we can write%
\begin{equation}
\left[ \Delta (0)\right] ^{\frac{l}{2}-1}=\left( \underset{j=-1}{\overset{%
+\infty }{\sum }}\alpha _{j}\epsilon ^{j}\right) ^{\frac{l}{2}-1}=\underset{%
j=-(\frac{l}{2}-1)}{\overset{+\infty }{\sum }}\xi _{j}^{(l)}\epsilon ^{j}=%
\frac{1}{\epsilon ^{\frac{l}{2}-1}}\underset{j=0}{\overset{+\infty }{\sum }}%
\xi _{j-(\frac{l}{2}-1)}^{(l)}\epsilon ^{j}  \label{new5}
\end{equation}%
where $\xi _{j-(\frac{l}{2}-1)}^{(l)}$ are some coefficients that depends on
the coefficients $\alpha _{j}$, for example%
\begin{gather}
\xi _{-(\frac{l}{2}-1)}^{(l)}=(\alpha _{-1})^{\frac{l}{2}-1}  \label{new5.1}
\\
\xi _{j-1}^{(4)}=\alpha _{j-1}\text{ \ \ \ for \ \ }j=0,1,2,...  \notag \\
\xi _{j-2}^{(6)}=\underset{k=0}{\overset{j}{\sum }}\alpha _{k-1}\alpha
_{j-k-1}\text{ }  \notag
\end{gather}

When we apply the dimensional regularization to make finite $\Delta (0)$, we
must introduce a mass factor $\mu $ to keep the coupling constant as a
dimensionless constant, this means that we have to replace $\lambda _{0}$
with $\lambda _{0}\mu ^{-\epsilon }$.\footnote{%
Because we are using $\hbar =c=1$ (god-given units), the dimension of $%
\lambda _{0}$ in a space-time of dimension $d$, is $\left[ mass\right]
^{d-4} $. But $\lambda _{0}$ is the "small" parameter to expand the
correlation function, then it has to be dimensionless, so we must introduce
some factor to cancel the mass-dependence behavior of $\lambda _{0}$. This
is achieved by changing $\lambda _{0}\rightarrow \lambda _{0}\mu ^{-\epsilon
}$ where $\left[ \mu \right] =\left[ mass\right] $ and $\epsilon =d-4$.}
Then eq.(\ref{new.2.1}) reads%
\begin{equation}
Tr(\rho _{int}^{(l,2,1)})=-\lambda _{0}\mu ^{-\epsilon }\left[ \Delta (0)%
\right] ^{\frac{l}{2}-1}  \label{new5.2}
\end{equation}%
We can expand $\mu ^{-\epsilon }$ as a Taylor series around $\epsilon =0$,
then the first order correction to the two-point correlation function reads%
\begin{equation}
Tr(\rho _{int}^{(l,2,1)})=-\frac{\lambda _{0}^{(l)}}{\epsilon ^{\frac{l}{2}%
-1}}\underset{k=0}{\overset{+\infty }{\sum }}\frac{(-1)^{k}(\ln \mu )^{k}}{k!%
}\epsilon ^{k}\left( \underset{j=0}{\overset{+\infty }{\sum }}\xi _{j-(\frac{%
l}{2}-1)}^{(l)}\epsilon ^{j}\right)  \label{new6}
\end{equation}%
Using that $\underset{n=0}{\sum }a_{n}\underset{n=0}{\sum }b_{n}=\underset{%
n=0}{\overset{+\infty }{\sum }}\overset{n}{\underset{k=0}{\sum }}%
a_{k}b_{n-k} $, then eq.(\ref{new6}) reads%
\begin{equation}
Tr(\rho _{int}^{(l,2,1)})=-\frac{\lambda _{0}^{(l)}}{\epsilon ^{\frac{l}{2}%
-1}}\underset{n=0}{\overset{+\infty }{\sum }}\epsilon ^{n}\overset{n}{%
\underset{k=0}{\sum }}\frac{(-1)^{k}(\ln \mu )^{k}}{k!}\xi _{n-k-(\frac{l}{2}%
-1)}^{(l)}  \label{new7}
\end{equation}%
The last sum is a Laurent expansion, where the principal part are the first $%
\frac{l}{2}-2$ terms. In the minimal subtraction scheme (see \cite{Collins},
page 56), this principal part is canceled with counterterms which do not
contain finite and arbitrary terms, then the renormalized result is the
coefficient of the last equation that multiplies to $\epsilon ^{0}$. In the
observable-state model, the projector operator $\Pi _{1}$ (see Section II.B
of \cite{PR2}) acting on $\rho _{int}^{(l,2,1)}$, gives a new quantum state $%
\Pi _{1}(\rho _{int}^{(l,2,1)})$. The trace of this projected quantum state
is the coefficient in eq.(\ref{new7}) that multiply to $\epsilon ^{0}$ which
is equivalent to the renormalized result of the minimal subtraction scheme.
Then%
\begin{equation}
Tr(\Pi _{1}(\rho _{int}^{(l,2,1)}))=\beta _{0}^{(l,2,1)}=\underset{i=1}{%
\overset{\frac{l}{2}-1}{\dprod }}\rho _{ND}^{(l,2,1,i)}=-\overset{\frac{l}{2}%
-1}{\underset{k=0}{\sum }}\frac{(-1)^{k}(\ln \mu )^{k}}{k!}\xi _{-k}^{(l)}
\label{new7.2}
\end{equation}%
where the second and third term of last equation comes from the
observable-state model (see eq.(24) or eq.(26) of \cite{PR2}). The product
of $\frac{l}{2}-1$ non-diagonal quantum states is the consequence of how the
internal quantum states $\rho _{int}$ are defined.\footnote{%
In Section 2.1.1 we show how this projector acts on the quantum state for
the $l=6$ case.}

\subsection{Mass renormalization group equation}

In \cite{PR2} we have shown how the renormalization group arises in the
context of the observable-state model for a $\phi ^{4}$ interaction. The
mass renormalization, when we take in account all the orders in the
perturbation expansion reads\footnote{%
In the following equations we will restore the Planck constant $\hbar $ for
later convenience.}\ 
\begin{equation}
m^{2}=m_{0}^{2}+\underset{p=1}{\overset{+\infty }{\sum }}(-1)^{p}\left(
\lambda _{0}^{(l)}\right) ^{p}\hbar ^{p}\beta _{0}^{(l,2,p)}(m_{0}^{2},\mu
)=m_{0}^{2}-\lambda _{0}^{(l)}\hbar \beta _{0}^{(l,2,1)}(m_{0}^{2},\mu )+...
\label{rg1}
\end{equation}%
In the other side, since $m_{0}^{2}$ and $\lambda _{0}$ do not depend on $%
\mu $ in the absence of loop correction, we have 
\begin{equation}
\frac{dm_{0}^{2}}{d\mu }=O(\hbar )\text{ \ \ \ \ \ \ \ }\frac{d\lambda
_{0}^{(l)}}{d\mu }=O(\hbar )  \label{rg2.2}
\end{equation}%
The renormalization group can be obtained by imposing the fact that the mass 
$m^{2}$ do not depend on $\mu $, this is, $\frac{dm^{2}}{d\mu }=0$. Using
the chain rule in eq.(\ref{rg1}), we have for $m^{2}$:%
\begin{equation}
\frac{dm^{2}}{d\mu }=\frac{\partial m^{2}}{\partial m_{0}^{2}}\frac{%
dm_{0}^{2}}{d\mu }+\frac{\partial m^{2}}{\partial \lambda _{0}^{(l)}}\frac{%
d\lambda _{0}^{(l)}}{d\mu }+\frac{\partial m^{2}}{\partial \mu }=0
\label{rg3}
\end{equation}%
using eqs.(\ref{rg1}) and (\ref{rg2.2}) at order $\hbar $, eq.(\ref{rg3})
reads 
\begin{equation}
\frac{dm_{0}^{2}}{d\mu }+\lambda _{0}^{(l)}\frac{\partial \beta
_{0}^{(l,2,1)}}{\partial \mu }=0  \label{rg6}
\end{equation}%
From eq.(\ref{new7.2}) we have that 
\begin{equation}
\beta _{0}^{(l,2,1)}=-\overset{\frac{l}{2}-1}{\underset{k=0}{\sum }}\frac{%
(-1)^{k}(\ln \mu )^{k}}{k!}\xi _{-k}^{(l)}  \label{rg7}
\end{equation}%
then%
\begin{equation}
\frac{\partial \beta _{0}^{(l,2,1)}}{\partial \mu }=-\frac{1}{\mu }\overset{%
\frac{l}{2}-2}{\underset{k=0}{\sum }}\frac{(-1)^{k+1}(\ln \mu )^{k}}{k!}\xi
_{-(k+1)}^{(l)}  \label{rg8}
\end{equation}%
replacing eq.(\ref{rg8}) in eq.(\ref{rg6}) we obtain a differential equation
for $m_{0}^{2}$ at order $\hbar $:%
\begin{equation}
\frac{dm_{0}^{2}}{d\mu }=\frac{\lambda _{0}^{(l)}}{\mu }\overset{\frac{l}{2}%
-2}{\underset{k=0}{\sum }}\frac{(-1)^{k+1}(\ln \mu )^{k}}{k!}\xi
_{-(k+1)}^{(l)}  \label{rg9}
\end{equation}%
we have shown that for $l=4$ (see \cite{PR2}), the differential equation
reads%
\begin{equation}
\frac{dm_{0}^{2}}{d\mu }=-\frac{\lambda _{0}}{\mu }\xi _{-1}^{(4)}=-\frac{%
\lambda _{0}}{\mu }\alpha _{-1}=\frac{\lambda _{0}}{\mu }\frac{m_{0}^{2}}{%
8\pi ^{2}}  \label{rg.9.2}
\end{equation}%
which can be solved to obtain%
\begin{equation}
m_{0}^{2}=m_{S}^{2}(\frac{\mu }{\mu _{S}})^{\frac{\lambda _{0}}{8\pi ^{2}}}
\label{rg10}
\end{equation}%
where $m_{S}^{2}$ is the value of the mass when $\mu =$ $\mu _{S}$ (see
eq.(76) of \cite{PR2}, which agrees with eq.(4.6.20) and eq.(4.6.22), page
142 of \cite{Ramond} at order $\hbar $).

Thus, eq.(\ref{rg9}) is the mass renormalization group equation for a $\phi
^{l}$ interaction. It is a non-linear differential equation for $l>4$,
because of the $m_{0}^{2}$ dependence in the $\xi _{-(k+1)}^{(l)}$
coefficients. For this reason, approximate methods must be used to solve it.
Is not the purpose of this work to obtain the exact relation between the
mass $m_{0}$ and the energy scale $\mu $, but only to obtain the behavior of
this relation for the $l=6$ case.

\subsubsection{First order in $\protect\phi ^{6}$ interaction for the $n=2$
correlation function}

In the case of $l=6$, the quantum state for the first order in the
perturbation expansion reads%
\begin{equation}
\rho _{ext}^{(6,2,1)}=\dint \left( \dint\limits_{{}}^{{}}\frac{d^{4}p}{(2\pi
)^{4}}\frac{e^{-ip(x_{1}-x_{2})}}{(p^{2}-m_{0}^{2})^{2}}\right) \left\vert
x_{1}\right\rangle \left\langle x_{2}\right\vert d^{4}x_{1}d^{4}x_{2}
\label{f1}
\end{equation}%
and%
\begin{gather}
\rho _{int}^{(6,2,1)}=\dint \left[ \rho _{D}^{(6,2,1,1)}(y_{1})\delta
(y_{1}-w_{1})+\rho _{ND}^{(6,2,1,1)}(y_{1},w_{1})\right]  \label{f2} \\
\left[ \rho _{D}^{(6,2,1,2)}(y_{2})\delta (y_{2}-w_{2})+\rho
_{ND}^{(6,2,1,2)}(y_{2},w_{2})\right] \left\vert y_{1},y_{2}\right\rangle
\left\langle w_{1},w_{2}\right\vert d^{4}y_{1}d^{4}y_{2}d^{4}w_{1}d^{4}w_{2}
\notag
\end{gather}%
where $\rho _{D}^{(6,2,1,i)}(y_{i})$ and $\rho
_{ND}^{(6,2,1,i)}(y_{i},w_{i}) $ for $i=1,2$ are analytical functions of its
arguments.\footnote{%
The observable-state model introduce a couple of diagonal quantum state $%
\rho _{D}$ and non-diagonal quantum state $\rho _{ND}$ for each loop in the
Feynman diagram, as it can be seen in \cite{PR2}, eq.(18).}

The observable reads%
\begin{equation}
O^{(6,2,1)}=\dint J(x_{1})J(x_{2})\left\vert x_{1},y_{1},y_{2}\right\rangle
\left\langle x_{2},y_{1},y_{2}\right\vert
d^{4}x_{1}d^{4}x_{2}d^{4}y_{1}d^{4}y_{2}  \label{f3}
\end{equation}%
where $J(x_{i})$ are the external sources that are introduced in the
generating functional.\footnote{%
These external sources can be considered as plane waves, where the Fourier
components are the positive and negative amount of energy of the ingoing and
outgoing particles. These values are the eigenvalues of one of the Casimir
operator $M^{2}=p^{\mu }p_{\mu }$ of the Poincare group, that is valid in
the asymptotic times $t\rightarrow \pm \infty $ (see \cite{Haag}, page 75).}

Because the observable contains an identity in the $y_{1}$ and $y_{2}$
coordinates, we must take the trace of $\rho _{int}^{(6,2,1)}$, then%
\begin{equation}
Tr(\rho _{int}^{(6,2,1)})=\rho _{D}^{(6,2,1,1)}\rho _{D}^{(6,2,1,2)}\epsilon
^{-2}+\left( \rho _{D}^{(6,2,1,1)}\rho _{ND}^{(6,2,1,2)}+\rho
_{D}^{(6,2,1,2)}\rho _{ND}^{(6,2,1,2)}\right) \epsilon ^{-1}+\rho
_{ND}^{(6,2,1,1)}\rho _{ND}^{(6,2,1,2)}  \label{f4}
\end{equation}%
where $\rho _{D}^{(6,2,1,1)}$, $\rho _{ND}^{(6,2,1,1)}$, $\rho
_{D}^{(6,2,1,2)}$ and $\rho _{ND}^{(6,2,1,2)}$ are normalization factors and
reads%
\begin{eqnarray}
\rho _{D}^{(6,2,1,1)} &=&\dint \rho _{D}^{(6,2,1,1)}(y_{1})d^{4}y_{1}\text{
\ \ \ \ \ \ \ \ \ \ \ \ }\rho _{D}^{(6,2,1,2)}=\dint \rho
_{D}^{(6,2,1,2)}(y_{2})d^{4}y_{2}  \label{f5} \\
\rho _{ND}^{(6,2,1,1)} &=&\dint \rho _{ND}^{(6,2,1,1)}(y_{1})d^{4}y_{1}\text{
\ \ \ \ \ \ \ \ \ \ \ \ }\rho _{ND}^{(6,2,1,2)}=\dint \rho
_{ND}^{(6,2,1,2)}(y_{2})d^{4}y_{2}  \notag
\end{eqnarray}%
Using eq.(\ref{new7.2})\ with $l=6$ we have%
\begin{eqnarray}
\rho _{D}^{(6,2,1,1)}\rho _{D}^{(6,2,1,2)} &=&\beta _{-2}^{(6,2,1)}=\xi
_{-2}^{(6)}  \label{f6} \\
\rho _{D}^{(6,2,1,1)}\rho _{ND}^{(6,2,1,2)}+\rho _{D}^{(6,2,1,2)}\rho
_{ND}^{(6,2,1,2)} &=&\beta _{-1}^{(6,2,1)}=\xi _{-1}^{(6)}-(\ln \mu )\xi
_{-2}^{(6)}  \notag \\
\rho _{ND}^{(6,2,1,1)}\rho _{ND}^{(6,2,1,2)} &=&\beta _{0}^{(6,2,1)}=\xi
_{0}^{(6)}-(\ln \mu )\xi _{-1}^{(6)}+\frac{(\ln \mu )^{2}}{2!}\xi _{-2}^{(6)}
\notag
\end{eqnarray}%
where, using eq.(\ref{new5}), the coefficients $\xi _{-2}^{(6)}$, $\xi
_{-1}^{(6)}$ and $\xi _{0}^{(6)}$ reads 
\begin{gather}
\xi _{-2}^{(6)}=(\alpha _{-1})^{2}=\frac{m_{0}^{4}}{64\pi ^{4}}  \label{f7}
\\
\xi _{-1}^{(6)}=2\alpha _{-1}\alpha _{0}=\frac{m_{0}^{4}}{64\pi ^{2}}\left(
\ln (\frac{m_{0}^{2}}{4\pi })+\gamma -1\right)  \notag \\
\xi _{0}^{(6)}=2\alpha _{-1}\alpha _{1}+\alpha _{0}^{2}=-\frac{m_{0}^{4}}{%
256\pi ^{4}}\left( \ln (\frac{m_{0}^{2}}{4\pi })\left( 2\gamma -1+\ln (\frac{%
m_{0}^{2}}{4\pi })\right) +\frac{\pi ^{2}}{6}+\gamma ^{2}-\gamma +1\right) 
\notag
\end{gather}%
The projector that gives the finite contribution to the first order in the
perturbation expansion reads%
\begin{gather}
\Pi _{1}(\rho _{int}^{(6,2,1)})=\rho _{int}^{(6,2,1)}-\dint \rho
_{D}^{(6.2,1,1)}(y_{1})\rho _{D}^{(6.2,1,2)}(y_{2})\left\vert
y_{1},y_{2}\right\rangle \left\langle y_{1},y_{2}\right\vert
d^{4}y_{1}d^{4}y_{2}-  \label{f8} \\
-\dint \rho _{D}^{(6.2,1,1)}(y_{1})\rho
_{ND}^{(6.2,1,2)}(y_{2},w_{2})\left\vert y_{1},y_{2}\right\rangle
\left\langle y_{1},w_{2}\right\vert d^{4}y_{1}d^{4}y_{2}d^{4}w_{2}-  \notag
\\
-\dint \rho _{ND}^{(6.2,1,1)}(y_{1},w_{1})\rho
_{D}^{(6.2,1,2)}(y_{2})\left\vert y_{1},y_{2}\right\rangle \left\langle
w_{1},y_{2}\right\vert d^{4}y_{1}d^{4}y_{2}d^{4}w_{1}  \notag
\end{gather}%
This means that we subtract from $\rho _{int}^{(6,2,1)}$ the terms that has,
at least, one diagonal quantum state. Then%
\begin{equation}
\Pi _{1}(\rho _{int}^{(6,2,1)})=\dint \rho
_{ND}^{(6,2,1,1)}(y_{1},w_{1})\rho _{ND}^{(6,2,1,2)}(y_{2},w_{2})\left\vert
y_{1},y_{2}\right\rangle \left\langle w_{1},w_{2}\right\vert
d^{4}y_{1}d^{4}y_{2}d^{4}w_{1}d^{4}w_{2}  \label{f9}
\end{equation}%
which is the finite contribution (see eq.(\ref{f6})). Then, if we take the
mean value of $O^{(6,2,1)}$ in the quantum state $\rho _{ext}^{(6,2,1)}\Pi
_{1}(\rho _{int}^{(6,2,1)})$ we have%
\begin{equation}
Tr(\rho _{ext}^{(6,2,1)}\Pi _{1}(\rho _{int}^{(6,2,1)})O^{(6,2,1)})=\beta
_{0}^{(6,2,1)}\dint \left( \dint\limits_{{}}^{{}}\frac{d^{4}p}{(2\pi )^{4}}%
\frac{e^{-ip(x_{1}-x_{2})}}{(p^{2}-m_{0}^{2})^{2}}\right)
J(x_{1})J(x_{2})d^{4}x_{1}d^{4}x_{2}  \label{f10}
\end{equation}

From eq.(\ref{f6})\ we can see that we have an indetermination for the
diagonal and non-diagonal quantum states. In fact, this indetermination
grows up as $L-1$, where $L~$is the number of loops at order $p$. This can
be obtained by noting that we have $2L$ unknown values, $L$ coming from the
diagonal quantum states and the other $L$ coming from the non-diagonal
quantum states, but, we only have $L+1$ equations (see eq.(23) of \cite{PR2}%
). From a different viewpoint, the indetermination can be obtained by noting
that the finite contribution, which is a unique equation (see eq.(\ref%
{new7.2})), is the product of $L$ non-diagonal quantum states. This point
deserves to be studied in more detail, since it is possible to apply a
unitary transformation to the diagonal and non-diagonal quantum states
resulting in a new diagonal and non-diagonal quantum states. But these new
quantum states must obey eq.(\ref{f6}) and this will introduce constraints
on the unitary transformation.\footnote{%
It is source of future works to determine if two succesive unitary
transformation obeys the constraint imposed in eq.(\ref{f6}), then the
unitary transformation would be the representation of some symmetry group.}

\subsubsection{The mass renormalization group for $\protect\phi ^{6}$}

In the case of $l=6$, from eq.(\ref{rg9})\ we obtain%
\begin{equation}
\frac{dm_{0}^{2}}{d\mu }=\lambda _{0}\frac{1}{\mu }\left( \xi
_{-1}^{(6)}-(\ln \mu )\xi _{-2}^{(6)}\right)  \label{rg5.2}
\end{equation}%
using eq.(\ref{f7}) we obtain%
\begin{equation}
\frac{dm_{0}^{2}}{d\mu }=\frac{\lambda _{0}}{\mu }\frac{m_{0}^{4}}{64\pi ^{4}%
}\left( \ln \left( \frac{4\pi \mu }{m_{0}^{2}}\right) +\gamma -1\right)
\label{rg5.3}
\end{equation}%
This differential equation is not linear and the dependent and independent
variables cannot be separated.

We can write eq.(\ref{rg5.3}) in more compact form by calling $f=m_{0}^{2}$, 
$\overline{\lambda }_{0}=\frac{\lambda _{0}}{64\pi ^{4}}$ and $\gamma -1=\ln
(c_{0})$, then

\begin{equation}
\frac{df}{d\mu }=-\overline{\lambda }_{0}\frac{f^{2}}{\mu }\ln \left( \frac{f%
}{a\mu }\right)  \label{rg5.3.1.1}
\end{equation}%
where $a=4\pi c_{0}$. Finally, we can make the following change of variables

\begin{equation}
z=\ln (f)\text{ \ \ \ \ \ \ \ }r=\ln (a\mu )  \label{rg5.3.1.1.0}
\end{equation}%
and eq.(\ref{rg5.3.1.1}) reads%
\begin{equation}
\frac{dz}{dr}=\overline{\lambda }_{0}e^{z}(r-z)  \label{rg5.3.1.1.1}
\end{equation}

Last equation is the most compact form in which we can write the relation
between the mass $m_{0}^{2}$ and the energy scale $\mu $ through the
definitions of eq.(\ref{rg5.3.1.1.0}).\footnote{%
The solution $z(r)$ must computed by numerical methods such as those
introduced in \cite{Birkhoff}. This will be source of a future work.} From
last equation, we can see the term $\overline{\lambda }_{0}e^{z}$ is
strictly positive, then the sign of the derivate $\frac{dz}{dr}$ depends on
the difference $r-z$. If $z(r)>r$ the function $z$ will decrease with $r$
and if $z(r)<r$ the function $z$ will increase with $r.$

\section{First correction to the coupling constant in $\protect\phi ^{l}$
theories}

To obtain the first correction of the coupling constant for a $\phi ^{l}$
theory, we must solve the second order in the perturbation expansion of the
correlation function of $l$-external points:%
\begin{equation}
G^{(l,l,2)}(x_{1},...,x_{l})=(-i\lambda _{0}\mu ^{-\epsilon })^{2}\dint
d^{4}y_{1}d^{4}y_{2}\left\langle \Omega _{0}\left\vert \phi (x_{1})...\phi
(x_{l})\phi ^{l}(y_{1})\phi ^{l}(y_{2})\right\vert \Omega _{0}\right\rangle
\label{landa1}
\end{equation}%
where the first and second superscript in $G$ refers to the power of the
interaction and the number of external points respectively and the $2$
superscript refers to the second order in the perturbation expansion. In
appendix A we show how to solve last equation which reads%
\begin{equation}
G^{(l,l,2)}(x_{1},...,x_{l})=-\lambda _{0}^{2}\left[ \underset{k=0}{\overset{%
\frac{l}{2}-1}{\sum }}\frac{(-2\ln \mu )^{k}}{k!}S_{\frac{l}{2}%
-1-k}^{(l)}+f_{0}^{(l)}\right] \underset{i=1}{\overset{l}{\dprod }}\dint 
\frac{d^{4}p_{i}}{(2\pi )^{4}}\frac{e^{-ip_{i}\cdot x_{i}}}{%
p_{i}^{2}-m_{0}^{2}}\delta (\overset{\frac{l}{2}}{\underset{j=1}{\dsum }}%
\left( p_{j}-p_{\frac{l}{2}+j}\right) )  \label{landa2}
\end{equation}%
where $S_{\frac{l}{2}-1-k}^{(l)}$ and $f_{0}^{(l)}$ are defined in eq.(\ref%
{apa17}) and eq.(\ref{apa14}) of Appendix A.

Then, using eq.(14) of \cite{PR2}, we can write%
\begin{equation}
\rho _{ext}^{(l,l,2)}=\underset{i=1}{\overset{l}{\dprod }}\dint \frac{%
d^{4}p_{i}}{(2\pi )^{4}}\frac{e^{-ip_{i}\cdot x_{i}}}{p_{i}^{2}-m_{0}^{2}}%
\delta (\overset{\frac{l}{2}}{\underset{j=1}{\dsum }}\left( p_{j}-p_{\frac{l%
}{2}+j}\right) )  \label{landa2.1}
\end{equation}%
and from eq.(\ref{apa20})%
\begin{equation}
Tr(\rho _{int}^{(l,l,2)})=\beta _{n-(\frac{l}{2}-1)}^{(l,l,2)}=-\frac{1}{%
\epsilon ^{\frac{l}{2}-2}}\underset{n=0}{\overset{+\infty }{\sum }}\epsilon
^{n-1}\underset{k=0}{\overset{n}{\sum }}\frac{(-2\ln \mu )^{k}}{k!}%
S_{n-k}^{(l)}+\underset{n=0}{\overset{+\infty }{\sum }}\epsilon ^{n}\underset%
{k=0}{\overset{n}{\sum }}\frac{(-2\ln \mu )^{k}}{k!}f_{n-k}^{(l)}
\label{landa4}
\end{equation}%
The finite value of $Tr(\rho _{int}^{(l,l,2)})$ is obtained by applying the
projector $\Pi _{2}(\rho _{int}^{(l,l,1)})$ introduced in eq.(24) or eq.(26)
of \cite{PR2}\thinspace ,\ which gives%
\begin{equation}
Tr(\Pi _{1}(\rho _{int}^{(l,l,2)}))=\beta _{0}^{(l,l,2)}=\underset{i=1}{%
\overset{\frac{l}{2}-1}{\dprod }}\rho _{ND}^{(l,2,1,i)}=\underset{k=0}{%
\overset{\frac{l}{2}-1}{\sum }}\frac{(-2\ln \mu )^{k}}{k!}S_{\frac{l}{2}%
-1-k}^{(l)}+f_{0}^{(l)}  \label{landa5}
\end{equation}%
and this correspond to the value $n=\frac{l}{2}-1$ in eq.(\ref{landa4}). In
the next section we will show how the projection procedure applies to a $%
\phi ^{6}$ interaction.

\subsection{Coupling constant renormalization group equation}

Following the same steps for the mass renormalization group equation, the
coupling constant renormalization, when we take in account all the orders in
the perturbation expansion, reads\ 
\begin{equation}
\lambda ^{(l)}=\lambda _{0}^{(l)}+\underset{p=2}{\overset{+\infty }{\sum }}%
\hbar ^{p-1}(-\lambda _{0}^{(l)})^{p}\beta _{0}^{(l,l,p)}(m_{0}^{2},\mu
)=\lambda _{0}^{(l)}+\hbar (\lambda _{0}^{(l)})^{2}\beta
_{0}^{(l,l,2)}(m_{0}^{2},\mu )+...  \label{coco1}
\end{equation}%
In the other side, since $\lambda _{0}^{(l)}$ and $m_{0}^{2}$ do not depend
on $\mu $ in the absence of loop correction, we have 
\begin{equation}
\frac{d\lambda _{0}^{(l)}}{d\mu }=O(\hbar )\text{ \ \ \ \ \ }\frac{dm_{0}^{2}%
}{d\mu }=O(\hbar )  \label{coco2}
\end{equation}%
The renormalization group can be obtained by imposing the fact that the
coupling constant $\lambda ^{(l)}$ do not depend on $\mu $, this is, $\frac{%
d\lambda ^{(l)}}{d\mu }=0$. Using the chain rule in eq.(\ref{coco1}), we
have for $\lambda ^{(l)}$:%
\begin{equation}
\frac{d\lambda ^{(l)}}{d\mu }=\frac{\partial \lambda ^{(l)}}{\partial
m_{0}^{2}}\frac{d\lambda _{0}^{(l)}}{d\mu }+\frac{\partial \lambda ^{(l)}}{%
\partial \lambda _{0}}\frac{d\lambda _{0}^{(l)}}{d\mu }+\frac{\partial
\lambda ^{(l)}}{\partial \mu }=0  \label{coco3}
\end{equation}%
using eqs.(\ref{coco1}) and (\ref{coco2}), eq.(\ref{coco3}) reads at order $%
\hbar $:%
\begin{equation}
\frac{d\lambda _{0}^{(l)}}{d\mu }+(\lambda _{0}^{(l)})^{2}\frac{\partial
\beta _{0}^{(l,l,1)}}{\partial \mu }=0  \label{coco4}
\end{equation}%
From eq.(\ref{landa5}) we have that 
\begin{equation}
\beta _{0}^{(l,l,2)}=\underset{k=0}{\overset{\frac{l}{2}-1}{\sum }}\frac{%
(-2\ln \mu )^{k}}{k!}S_{\frac{l}{2}-1-k}^{(l)}+f_{0}^{(l)}  \label{coco5}
\end{equation}%
then%
\begin{equation}
\frac{\partial \beta _{0}^{(l,l,2)}}{\partial \mu }=\frac{1}{\mu }\underset{%
k=0}{\overset{\frac{l}{2}-2}{\sum }}\frac{(-2\ln \mu )^{k}}{k!}S_{\frac{l}{2}%
-2-k}^{(l)}  \label{coco6}
\end{equation}%
replacing eq.(\ref{coco6}) in eq.(\ref{coco4}) we obtain a differential
equation for $\lambda _{0}^{(l)}$ at order $\hbar $:%
\begin{equation}
\frac{d\lambda _{0}^{(l)}}{d\mu }=-(\lambda _{0}^{(l)})^{2}\frac{1}{\mu }%
\underset{k=0}{\overset{\frac{l}{2}-2}{\sum }}\frac{(-2\ln \mu )^{k}}{k!}S_{%
\frac{l}{2}-2-k}^{(l)}  \label{coco7}
\end{equation}%
which can be solved by separating $\lambda _{0}^{(l)}$ and $\mu $ in each
sides of last equation:%
\begin{equation}
\dint_{\lambda _{S}^{(l)}}^{\lambda _{0}^{(l)}}\frac{d\lambda _{0}^{\prime
(l)}}{(\lambda _{0}^{\prime (l)})^{2}}=-\underset{k=0}{\overset{\frac{l}{2}-2%
}{\sum }}\frac{S_{\frac{l}{2}-2-k}^{(l)}}{k!}\dint_{\mu _{S}}^{\mu }\frac{%
d\mu ^{\prime }}{\mu ^{\prime }}(-2\ln \mu ^{\prime })^{k}  \label{coco8}
\end{equation}%
and using that%
\begin{equation}
\dint\limits_{{}}^{{}}\frac{dx}{x}(-2\ln x)^{n}=-\frac{1}{2(n+1)}(-2\ln
x)^{n+1}  \label{coco9}
\end{equation}%
eq.(\ref{coco8})\ reads%
\begin{equation}
\lambda _{0}^{(l)}=\frac{\lambda _{S}^{(l)}}{1+\lambda _{S}^{(l)}\underset{%
k=0}{\overset{\frac{l}{2}-2}{\sum }}\Lambda _{k}^{(l)}\left( \ln ^{k+1}\mu
-\ln ^{k+1}\mu _{S}\right) }  \label{coco10}
\end{equation}%
where $\lambda _{0}^{(l)}(\lambda _{S}^{(l)})=\mu _{S}$ and 
\begin{equation}
\Lambda _{k}^{(l)}=\frac{S_{\frac{l}{2}-2-k}^{(l)}}{(k+1)!}(-2)^{k}
\label{coco11}
\end{equation}

For example, for $l=4$ we obtain%
\begin{equation}
\lambda _{0}^{(4)}=\frac{\lambda _{S}^{(4)}}{1+\lambda _{S}^{(4)}\Lambda
_{0}^{(4)}\ln \left( \frac{\mu }{\mu _{S}}\right) }  \label{coco12}
\end{equation}%
where (see eq.(\ref{apa17})) of Appendix A%
\begin{equation}
\Lambda _{0}^{(4)}=S_{0}^{(4)}=-\frac{3}{16\pi ^{2}}  \label{coco13}
\end{equation}%
then eq.(\ref{coco12}) is identical to eq.(4.6.16) of \cite{Ramond}.

\paragraph{Domain of validity of the perturbation expansion}

From eq.(\ref{coco12}) we can see that $\lambda _{0}^{(4)}$ increases with $%
\mu $, but $\lambda _{0}^{(4)}$ is the parameter that we use to expand the
perturbation of the correlation functions, so we must demand that this
parameter do not leaves the domain of validity of the perturbation theory,
that is, $\left\vert \lambda _{0}^{(4)}\right\vert <1$. But using eq.(\ref%
{coco12}), this implies that%
\begin{equation}
\mu <e^{\frac{16\pi ^{2}}{3\lambda _{S}^{(4)}}(1-\lambda _{S}^{(4)})}
\label{coco14}
\end{equation}%
where we put $\mu _{S}=1$ without loss of generality. Restoring the Planck
constant $\hbar $ and the velocity of light $c$, the last inequality can be
written in terms of a characteristic distance $d=\frac{\hbar }{\mu c}$.
Then, the last inequality reads%
\begin{equation}
d>\frac{\hbar }{c}e^{-\frac{16\pi ^{2}}{3\lambda _{S}^{(4)}}(1-\lambda
_{S}^{(4)})}  \label{coco15}
\end{equation}

This means that interactions that occur at a distance less than $\frac{\hbar 
}{c}e^{-\frac{16\pi ^{2}}{3\lambda _{S}^{(4)}}(1-\lambda _{S}^{(4)})}$ are
out of scope of the perturbation expansion (see \cite{Ramond}, page 139).
This point deserves a more detailed study from the conceptual and
mathematical viewpoint because it is an argument for the projection method,
in fact, the projector remove the diagonal part of the quantum state, that
is, the interaction at the point, but eq.(\ref{coco15})\ implies that we
have to remove the interactions occurring below $d$. In this sense, the
projection procedure is an approximation of an exact projection that
neglects the $\lambda _{0}^{(4)}$ sector that is out of the domain of
validity.\footnote{%
In this sense, the cut-off is not defined by a possible quantum gravity
theory, but rather it is defined as the value at which the coupling constant
is no longer a valid parameter for the perturbation expansion.} Also, this
suggest to write the correlation function as a Laurent series which
converges for $d<$ $\left\vert z\right\vert <R$. Following the same argument
on eq. (\ref{coco10}), we can see that the coupling constant $\lambda
_{0}^{(l)}$ depends with $\mu $, then the perturbation theory will give
reasonable results only when $\left\vert \lambda _{0}^{(l)}\right\vert <1$,
which implies that%
\begin{equation}
\frac{1}{\lambda _{S}^{(l)}}+\underset{k=0}{\overset{\frac{l}{2}-2}{\sum }}%
\Lambda _{k}^{(l)}x^{k+1}>1  \label{dom1}
\end{equation}%
where again we use that $\mu _{s}=1$ and $x=\ln \mu $.

Eq.(\ref{dom1}) is a polynomial inequality of order $\frac{l}{2}-1$, then
there will be, at least, $\frac{l}{2}-1$ inequalities for the maximum or
minimum distance where the interaction process can occur.

Summing up, the dimensional regularization introduces an arbitrary energy
scale or equivalently, an arbitrary distance scale. The perturbation
expansion puts a bound on the possible values of the coupling constant which
shows an upper or lower limit for the distance scale. Finally, we have to
remove the interaction that occurs below the lower limit or above the upper
limit, because the coupling constant is out of the domain of validity of the
perturbation expansion. The observable-state model uses the fact that if we
do not have a theory for the short-distance scale\footnote{%
This means, the expansion do not converge for some values of the expansion
parameter.}, we can remove it by a suitable projector operator, but this
projector also renormalize the theory, in the sense that takes away the
divergences that arise from the short-distance scale. In this sense, the
idea is not different from the Wilson approach of renormalization group \cite%
{Wilson}, where the high momentum modes are integrated out.

\subsubsection{First order in $\protect\phi ^{6}$ interaction for the $n=6$
correlation function}

In the case of $l=6$, the quantum state for the first order in the
perturbation expansion reads%
\begin{equation}
\rho _{ext}^{(6,6,2)}=\dint \left( \underset{i=1}{\overset{6}{\dprod }}%
\dint\limits_{{}}^{{}}\frac{d^{4}p_{i}}{(2\pi )^{4}}\frac{e^{-ip_{i}x_{i}}}{%
p_{i}^{2}-m_{0}^{2}}\delta (\overset{6}{\underset{j=1}{\dsum }}p_{j})\right)
\left\vert x_{1},x_{2},x_{3}\right\rangle \left\langle
x_{4,}x_{5},x_{6}\right\vert \underset{i=1}{\overset{6}{\dprod }}d^{4}x_{i}
\label{fi1}
\end{equation}%
where the first superscript $6$ refers to the power of interaction, the
second superscript $6$ refers to the number of external point of \ the
correlation function and the third superscript $2$ refers to the second
order in the perturbation expansion. The internal quantum state reads 
\begin{gather}
\rho _{int}^{(6,6,2)}=\dint \left[ \rho _{D}^{(6,6,2,1)}(y_{1})\delta
(y_{1}-w_{1})+\rho _{ND}^{(6,6,2,1)}(y_{1},w_{1})\right]  \label{fi2} \\
\left[ \rho _{D}^{(6,6,2,2)}(y_{2})\delta (y_{2}-w_{2})+\rho
_{ND}^{(6,6,2,2)}(y_{2},w_{2})\right] \left\vert y_{1},y_{2}\right\rangle
\left\langle w_{1},w_{2}\right\vert d^{4}y_{1}d^{4}y_{2}d^{4}w_{1}d^{4}w_{2}
\notag
\end{gather}%
and the observable reads%
\begin{equation}
O^{(6,6,2)}=\dint \underset{i=1}{\overset{6}{\dprod }}J(x_{i})\left\vert
x_{1},x_{2},x_{3}\right\rangle \left\langle x_{4,}x_{5},x_{6}\right\vert 
\underset{i=1}{\overset{6}{\dprod }}d^{4}x_{i}  \label{fi3}
\end{equation}%
Because the observable contains an identity in the $y_{1}$ and $y_{2}$
coordinates, we must take the trace of $\rho _{int}^{(6,6,2)}$, then%
\begin{equation}
Tr(\rho _{int}^{(6,6,2)})=\rho _{D}^{(6,6,2,1)}\rho _{D}^{(6,6,2,2)}\epsilon
^{-2}+\left( \rho _{D}^{(6,6,2,1)}\rho _{ND}^{(6,6,2,2)}+\rho
_{D}^{(6,6,2,2)}\rho _{ND}^{(6,6,2,1)}\right) \epsilon ^{-1}+\rho
_{ND}^{(6,6,2,1)}\rho _{ND}^{(6,6,2,2)}  \label{fi4}
\end{equation}%
where%
\begin{eqnarray}
\rho _{D}^{(6,6,2,1)} &=&\dint \rho _{D}^{(6,6,2,1)}(y_{1})d^{4}y_{1}\text{
\ \ \ \ \ \ \ \ \ \ \ \ }\rho _{D}^{(6,6,2,2)}=\dint \rho
_{D}^{(6,6,2,2)}(y_{2})d^{4}y_{2}  \label{fi5} \\
\rho _{ND}^{(6,6,2,1)} &=&\dint \rho _{ND}^{(6,6,2,1)}(y_{1})d^{4}y_{1}\text{
\ \ \ \ \ \ \ \ \ \ \ \ }\rho _{ND}^{(6,6,2,2)}=\dint \rho
_{ND}^{(6,6,2,2)}(y_{2})d^{4}y_{2}  \notag
\end{eqnarray}%
Using eq.(\ref{apa20}) of Appendix A\ with $l=6$ we have%
\begin{equation}
\Gamma ^{(6)}(p_{1},p_{2},p_{3})=S_{0}^{(6)}\epsilon ^{-2}+\left(
S_{1}^{(6)}-2\ln \mu S_{0}^{(6)}\right) \epsilon ^{-1}+S_{2}^{(6)}-2\ln \mu
S_{1}^{(6)}+2\ln ^{2}\mu S_{0}^{(6)}+f_{0}^{(6)}+O(\epsilon )  \label{fi6}
\end{equation}%
then%
\begin{gather}
\rho _{D}^{(6,6,2,1)}\rho _{D}^{(6,6,2,2)}=\beta _{-2}^{(6,6,2)}=S_{0}^{(6)}
\label{fi7} \\
\rho _{D}^{(6,6,2,1)}\rho _{ND}^{(6,6,2,2)}+\rho _{D}^{(6,6,2,2)}\rho
_{ND}^{(6,6,2,1)}=\beta _{-1}^{(6,6,2)}=S_{1}^{(6)}-2\ln \mu S_{0}^{(6)} 
\notag \\
\rho _{ND}^{(6,6,2,1)}\rho _{ND}^{(6,6,2,2)}=\beta
_{0}^{(6,6,2)}=S_{2}^{(6)}-2\ln \mu S_{1}^{(6)}+2\ln ^{2}\mu
S_{0}^{(6)}+f_{0}^{(6)}  \notag
\end{gather}%
where%
\begin{gather}
S_{0}^{(6)}=\frac{3m_{0}^{2}}{128\pi ^{4}}  \label{fi8} \\
S_{1}^{(6)}=-\frac{3m_{0}^{2}}{256\pi ^{4}}\left[ \gamma +1-\ln (\frac{4\pi 
}{m_{0}^{2}})\right]  \notag \\
S_{2}^{(6)}=-\frac{m_{0}^{2}}{8\pi ^{2}}\{\frac{1}{768}+\frac{1}{32\pi ^{2}}(%
\frac{\gamma }{2}-\ln (2\sqrt{\pi }))^{2}+\frac{3}{32\pi ^{2}}(1-\gamma +\ln
(\frac{4\pi }{m_{0}^{2}}))(\ln (\frac{4\pi }{m_{0}^{2}})-\gamma )+  \notag \\
\frac{3}{768\pi ^{2}}[6\ln (\frac{m_{0}^{2}}{4\pi })(2\gamma -2+\ln (\frac{%
m_{0}^{2}}{4\pi }))+\pi ^{2}+6\gamma ^{2}-12\gamma +12]\}  \notag
\end{gather}%
The projector that gives the finite contribution to the first order in the
perturbation expansion reads%
\begin{gather}
\Pi _{2}(\rho _{int}^{(6,6,2)})=\rho _{int}^{(6,6,2)}-\dint \rho
_{D}^{(6.6,2,1)}(y_{1})\rho _{D}^{(6.6,2,2)}(y_{2})\left\vert
y_{1},y_{2}\right\rangle \left\langle y_{1},y_{2}\right\vert
d^{4}y_{1}d^{4}y_{2}-  \label{fi9} \\
-\dint \rho _{D}^{(6.6,2,1)}(y_{1})\rho
_{ND}^{(6.6,2,2)}(y_{2},w_{2})\left\vert y_{1},y_{2}\right\rangle
\left\langle y_{1},w_{2}\right\vert d^{4}y_{1}d^{4}y_{2}d^{4}w_{2}-  \notag
\\
-\dint \rho _{ND}^{(6.6,2,1)}(y_{1},w_{1})\rho
_{D}^{(6.6,2,2)}(y_{2})\left\vert y_{1},y_{2}\right\rangle \left\langle
w_{1},y_{2}\right\vert d^{4}y_{1}d^{4}y_{2}d^{4}w_{1}  \notag
\end{gather}%
then%
\begin{equation}
\Pi _{2}(\rho _{int}^{(6,6,2)})=\dint \rho
_{ND}^{(6,6,2,1)}(y_{1},w_{1})\rho _{ND}^{(6,6,2,2)}(y_{2},w_{2})\left\vert
y_{1},y_{2}\right\rangle \left\langle w_{1},w_{2}\right\vert
d^{4}y_{1}d^{4}y_{2}d^{4}w_{1}d^{4}w_{2}  \label{fi10}
\end{equation}

Finally, the mean value of observable of eq.(\ref{fi3}) in the projected
quantum state reads

\begin{equation}
Tr(\rho _{ext}^{(6,2,2)}\Pi _{1}(\rho _{int}^{(6,6,2)})O^{(6,2,2)})=\beta
_{0}^{(6,2,2)}\dint \left( \underset{i=1}{\overset{6}{\dprod }}%
\dint\limits_{{}}^{{}}\frac{d^{4}p_{i}}{(2\pi )^{4}}\frac{e^{-ip_{i}x_{i}}}{%
p_{i}^{2}-m_{0}^{2}}\delta (\overset{6}{\underset{j=1}{\dsum }}p_{j})\right) 
\underset{k=1}{\overset{6}{\dprod }}J(x_{k})d^{4}x_{k}  \label{fi11}
\end{equation}

where, using eq.(\ref{fi7}), eq.(\ref{fi8}) and eq.(\ref{apa14}), $\beta
_{0}^{(6,2,2)}$ reads

\begin{gather}
\beta _{0}^{(6,2,2)}=-\frac{m_{0}^{2}}{8\pi ^{2}}\{\frac{1}{768}+\frac{1}{%
32\pi ^{2}}(\frac{\gamma }{2}-\ln (2\sqrt{\pi }))^{2}+\frac{3}{32\pi ^{2}}%
(1-\gamma +\ln (\frac{4\pi }{m_{0}^{2}}))(\ln (\frac{4\pi }{m_{0}^{2}}%
)-\gamma )+  \label{fi12} \\
\frac{3}{768\pi ^{2}}[6\ln (\frac{m_{0}^{2}}{4\pi })(2\gamma -2+\ln (\frac{%
m_{0}^{2}}{4\pi }))+\pi ^{2}+6\gamma ^{2}-12\gamma +12]\}+\frac{3m_{0}^{2}}{%
128\pi ^{4}}\left[ \gamma +1-\ln (\frac{4\pi }{m_{0}^{2}})\right] \ln \mu + 
\notag \\
\frac{3m_{0}^{2}}{64\pi ^{4}}\ln ^{2}\mu +f_{0}^{(6)}(p_{1},p_{2},p_{3}) 
\notag
\end{gather}%
where, using eq.(\ref{apa8}), eq.(\ref{apa10}) and eq.(\ref{apa14}), $%
f_{0}^{(6)}$ reads

\begin{equation}
f_{0}^{(6)}(p_{1},p_{2},p_{3})=\dint \frac{d^{4}q_{1}}{(2\pi )^{4}}\frac{1}{%
\left( q_{1}^{2}-m_{0}^{2}\right) }\left[ \frac{1}{32\pi ^{2}}\underset{%
z=s,t,u}{\overset{}{\sum }}\sqrt{1+\frac{4m_{0}^{2}}{z^{2}}}\ln \left\{ 
\frac{\sqrt{1+\frac{4m_{0}^{2}}{z^{2}}}+1}{\sqrt{1+\frac{4m_{0}^{2}}{z^{2}}}%
-1}\right\} \right]  \label{fi12.1}
\end{equation}%
where $s$, $t$ and $u$ are the Mandelstam variables 
\begin{equation}
s=(p_{1}+p_{2})^{2}\text{ \ \ \ \ \ \ \ \ \ }t=(p_{1}+p_{3})^{2}\text{ \ \ \
\ \ \ \ \ \ }u=(p_{1}+p_{4})^{2}  \label{fi12.2}
\end{equation}%
that we have to add in the evaluation of the six-point correlation function.

\subsubsection{The coupling constant renormalization group for $\protect\phi %
^{6}$}

In the case of $l=6$, from eq.(\ref{coco10})\ we obtain%
\begin{equation}
\lambda _{0}^{(6)}=\frac{1}{\frac{1}{\lambda _{S}^{(6)}}+\Lambda
_{0}^{(6)}\ln \mu +\Lambda _{1}^{(6)}\ln ^{2}\mu }  \label{fi13}
\end{equation}%
where we use that $\mu _{S}=1$ and using eq.(\ref{coco11}) and eq.(\ref%
{apa17}) 
\begin{equation}
\Lambda _{0}^{(6)}=S_{1}^{(6)}=\underset{k=0}{\overset{1}{\sum }}\xi
_{k-1}^{(4)}\eta _{-k}=\xi _{-1}^{(4)}\eta _{0}+\xi _{0}^{(4)}\eta
_{-1}=\alpha _{-1}\eta _{0}+\alpha _{0}\eta _{-1}=\frac{3m_{0}^{2}}{256\pi
^{4}}\left( \ln (\frac{4\pi }{m_{0}^{2}})-\gamma -1\right)  \label{fi14}
\end{equation}%
and%
\begin{equation}
\Lambda _{1}^{(6)}=-S_{0}^{(6)}=-\xi _{-1}^{(4)}\eta _{-1}=-\alpha _{-1}\eta
_{-1}=-\frac{3m_{0}^{2}}{128\pi ^{4}}  \label{fi15}
\end{equation}%
Then, $\Lambda _{0}^{(6)}$ can be either positive or negative and$~\Lambda
_{1}^{(6)}$ is strictly negative. Using eq.(\ref{dom1}), the domain of
validity of the coupling constant $\lambda _{0}^{(6)}$ is given by the
following inequality%
\begin{equation}
\frac{1}{\lambda _{S}^{(6)}}+\Lambda _{0}^{(6)}x-\left\vert \Lambda
_{1}^{(6)}\right\vert x^{2}>1  \label{fi16}
\end{equation}%
where $\ln \mu =x$. This last equation can be written as%
\begin{equation}
\left\vert x-\frac{\Lambda _{0}^{(6)}}{2\left\vert \Lambda
_{1}^{(6)}\right\vert }\right\vert <\sqrt{\left( \frac{\Lambda _{0}^{(6)}}{%
2\left\vert \Lambda _{1}^{(6)}\right\vert }\right) ^{2}+\frac{1}{\left\vert
\Lambda _{1}^{(6)}\right\vert \lambda _{S}^{(6)}}\left( 1-\lambda
_{S}^{(6)}\right) }  \label{fi16.1}
\end{equation}%
The two solutions are 
\begin{equation}
\mu _{1}<e^{q+v}  \label{fi16.2}
\end{equation}%
and%
\begin{equation}
\mu _{2}>e^{-q+v}  \label{fi16.3}
\end{equation}%
where 
\begin{equation}
q=\sqrt{\left( \frac{\Lambda _{0}^{(6)}}{2\left\vert \Lambda
_{1}^{(6)}\right\vert }\right) ^{2}+\frac{1}{\left\vert \Lambda
_{1}^{(6)}\right\vert \lambda _{S}^{(6)}}\left( 1-\lambda _{S}^{(6)}\right) }
\label{fi16.4}
\end{equation}%
and%
\begin{equation}
v=\frac{\Lambda _{0}^{(6)}}{2\left\vert \Lambda _{1}^{(6)}\right\vert }
\label{fi16.4.1}
\end{equation}%
Then, if $e^{2q}<1$, perturbation expansion converges in the region $\mu \in
\left( 0,e^{q+v}\right) \cup \left( e^{-q+v},+\infty \right) $ and if $%
e^{2q}>1$, the perturbation expansion converges in the region $\mu \in
\left( e^{-q+v},e^{q+v}\right) $. \ In terms of a distance scale, conditions
(\ref{fi16.3}) and (\ref{fi16.4})\ reads

\begin{equation}
d_{1}>e^{-q-v}  \label{fi16.5}
\end{equation}%
and%
\begin{equation}
d_{2}<e^{q-v}  \label{fi16.6}
\end{equation}

Then, if $e^{2q}<1$, perturbation expansion converges in the region $d\in
\left( 0,e^{q-v}\right) \cup \left( e^{-q-v},+\infty \right) $ and if $%
e^{2q}>1$, the perturbation expansion converges in the region $d\in
(e^{-q-v},e^{q-v})$. This first convergence region is not allowed because
the condition $e^{2q}<1$ implies that $q<0$, and there is no real number
whose square root is negative. This make sense because otherwise the
perturbation expansion would converge for distances between $0$ and $e^{q-v}$%
, but this would contradict the projector method, that remove the
short-distance interactions. The second convergence region implies that we
have to disregard the short-distance physics below $e^{-q-v}$ and above $%
e^{q-v}$ as occurs in Laurent series.\footnote{%
From eq.(\ref{fi16}), the coordinate of the vertex of the parabola is
located at $\left( v,\left\vert \Lambda _{1}^{(6)}\right\vert v^{2}+\frac{1}{%
\lambda _{S}^{(6)}}\right) $ and the parabola open downward. Then, if $%
\left\vert \Lambda _{1}^{(6)}\right\vert v^{2}+\frac{1}{\lambda _{S}^{(6)}}%
>1 $ we will have convergence for some $\mu $ points, but if $\left\vert
\Lambda _{1}^{(6)}\right\vert v^{2}+\frac{1}{\lambda _{S}^{(6)}}<1$ the
parabola will not cross the $y=1$ horizontal axis and we will not have
convergence points for the perturbation expansion.}

\section{Renormalization conditions}

In the observable-state model, the renormalization procedure is exchanged
for a projection operation acting on a Hilbert space. This implies that is
not necessary to introduce counterterms in the Lagrangian, but only to
remove the diagonal elements of multivalued distributions. From this
viewpoint, the observable-state model is only a formal development of the
ideas introduced in Section 3.7, page 59 of \cite{Collins}, "...the
divergences comes from a region in coordinate space where several
interactions occur very close to each other. The divergence can then be
cancelled by a counterterm which is a $\delta $-function in the position of
these interactions". In our formalism, this $\delta $-functions are
introduced as diagonal generalized functions and not as counterterms, which
are then discarded by the projector. Then the renormalization, from the
observable-state model viewpoint, is a procedure on the level of correlation
functions and not over the Lagrangian.

But then, is necessary to understand what are the $m_{0}$ and $m$ masses. In
this sense, we take the definition of the mass of a particle as the the
value of the momentum $p$ at which the propagator in momentum space has a
pole. If the propagator reads%
\begin{equation}
G_{0}(p)=\frac{i}{p^{2}-m_{0}^{2}}  \label{rc1}
\end{equation}%
then $p=m_{0}$ is the mass of the particle. In a similar way, the propagator
of the interacting theory reads%
\begin{equation}
G(p)=\frac{i}{p^{2}-(m_{0}^{2}+M^{2})}  \label{rc2}
\end{equation}%
where $M^{2}$ is the contribution of the one-particle irreducible Feynman
diagrams and without renormalization it reads%
\begin{equation}
M^{2}(m_{0},\mu ,\lambda _{0},p)=\underset{p=1}{\overset{+\infty }{\sum }}%
(-i\lambda _{0})^{p}\underset{j=-1}{\overset{+\infty }{\sum }}\beta
_{j}^{(2,p)}(m_{0},\mu ,p)\epsilon ^{j}  \label{rc3}
\end{equation}%
where $\beta _{j}^{(2,p)}$ are functions of the mass $m_{0}$ introduced in
the Lagrangian, $\mu $ is the energy scale introduced to maintain the
coupling constant $\lambda _{0}$ dimensionless and $p$ is the external
momentum (see eq.(B19) of \cite{PR1}). The projection leaves only the $j=0$
term in the second sum on eq.(\ref{rc3}), then%
\begin{equation}
M_{\text{finite}}^{2}(m_{0},\mu ,\lambda _{0},p)=\underset{p=1}{\overset{%
+\infty }{\sum }}(-i\lambda _{0})^{p}\beta _{0}^{(2,p)}(m_{0},\mu ,p)
\label{rc4}
\end{equation}%
Then, the mass of the self-interacting field is located where the
denominator of eq.(\ref{rc2})\ is zero, this is at the value $p^{2}=m^{2}$,
precisely

\begin{equation}
m^{2}-m_{0}^{2}-M_{\text{finite}}^{2}(m_{0}^{2},\mu ,m^{2},\lambda _{0})=0
\label{rc5}
\end{equation}

But in our renormalization scheme, we do not have counterterms, and in
particular, we do not have introduce a finite and arbitrary constant that
has no pole in $\epsilon =d-4$. Then, we cannot put $m^{2}=m_{0}^{2}$ in eq.(%
\ref{rc5}), because this implies that $M_{\text{finite}}^{2}(m_{0}^{2},\mu
,\lambda _{0})=0$, but the correction at first order do not depends on $p$,
so, there are no other term in $M_{\text{finite}}^{2}$ that cancel this
correction.\footnote{%
In fact, the following terms are multiplied with $(\lambda _{0})^{j}$, with $%
j>1$, where $j$ is the number of internal vertices, then, there is no term
independient of $p$ that multiplies $\lambda _{0}$, that cancel the first
correction of the two-point correlation function.} This means that $M_{\text{%
finite}}^{2}\neq 0$, so the renormalization condition introduced in page
328, eq.(10.28) of \cite{PS} cannot be applied in the observable-state
model. \footnote{%
This condition would implies that the propagator of the free field and the
interacting field are the same, which reminds us the Haag theorem (see \cite%
{Roman}, chapter 8).} But eq.(\ref{rc5})\ has two arbitrary parameters, $%
m^{2}$, which is the physical mass and $\mu $, which is an arbitrary mass
factor. Then, the mass renormalization group equation of eq. (\ref{rg3}) can
be obtained from the condition of eq.(\ref{rc5}) if we allow the mass $%
m_{0}^{2}$ introduced in the Lagrangian to be a function of $\mu $ and that
the physical mass do not depend on it.

\section{Conclusions}

The aim of this work was to extend the observable-state model to $\phi ^{l}$
self-interaction. We have found the first correction for the two-point
correlation function and the second correction to the $l-$correlation
function.\ Besides this, we have computed the mass and coupling
renormalization group equations at one loop correction. In the latter case
we have solved the differential equation finding the distance scale at which
the coupling constant leaves the domain of validity for the perturbation
expansion. This is an important result because it validates the projection
procedure.

\section{Acknowledgment}

This paper was partially supported by grants of CONICET (Argentina National
Research Council), FONCYT (Argentina Agency for Science and Technology) and
Universidad Nacional del Sur (UNS).

\appendix

\section{$l$-correlation function for a $\protect\phi ^{l}$ interaction}

Let us remember eq.(\ref{landa1}):%
\begin{equation}
G^{(l,l,2)}(x_{1},...,x_{l})=(-i\lambda _{0})^{2}\mu ^{-2\epsilon }\dint
d^{4}y_{1}d^{4}y_{2}\left\langle \Omega _{0}\left\vert \phi (x_{1})...\phi
(x_{l})\phi ^{4}(y_{1})\phi ^{4}(y_{2})\right\vert \Omega _{0}\right\rangle
\label{apa1}
\end{equation}%
the only connected diagram which contributes to $G^{(l,l,2)}$ reads%
\begin{equation}
G^{(l,l,2)}(x_{1},...,x_{l})=(-i\lambda _{0})^{2}\mu ^{-2\epsilon }\dint
d^{4}y_{1}d^{4}y_{2}[\Delta (y_{1}-y_{2})]^{\frac{l}{2}}\underset{k=1}{%
\overset{\frac{l}{2}}{\dprod }}\Delta (x_{k}-y_{1})\Delta (x_{\frac{l}{2}%
+k}-y_{2})  \label{apa2}
\end{equation}%
using that 
\begin{equation}
\Delta (x-y)=\dint\limits_{{}}^{{}}\frac{d^{4}p}{(2\pi )^{4}}\frac{%
e^{-ip(x-y)}}{p^{2}-m_{0}^{2}}  \label{apa3}
\end{equation}%
and integrating in $d^{4}y_{1}$ and $d^{4}y_{2}$, eq.(\ref{apa2}) reads%
\begin{equation}
G^{(l,l,2)}(x_{1},...,x_{l})=(-i\lambda _{0})^{2}\underset{i=1}{\overset{l}{%
\dprod }}\dint \frac{d^{4}p_{i}}{(2\pi )^{4}}\frac{e^{-ip_{i}\cdot x_{i}}}{%
p_{i}^{2}-m_{0}^{2}}\delta (\overset{\frac{l}{2}}{\underset{j=1}{\dsum }}%
\left( p_{j}-p_{\frac{l}{2}+j}\right) )\Gamma ^{(l)}(p_{1},...,p_{\frac{l}{2}%
})  \label{apa4}
\end{equation}%
where%
\begin{equation}
\Gamma ^{(l)}(p_{1},...,p_{\frac{l}{2}})=\mu ^{-2\epsilon }\underset{s=1}{%
\overset{\frac{l}{2}}{\dprod }}\dint \frac{d^{4}q_{s}}{(2\pi )^{4}}\frac{1}{%
q_{s}^{2}-m_{0}^{2}}\delta (\overset{\frac{l}{2}}{\underset{k=1}{\dsum }}%
\left( p_{k}-q_{k}\right) )  \label{apa5}
\end{equation}%
and where $p_{i}$ are the external momentum and $q_{i}$ are the loop
momentum. The two Dirac deltas are the conservation of momentum in each
vertices.

Using the Dirac delta with $q_{\frac{l}{2}}$ coordinate in eq.(\ref{apa5})
we obtain 
\begin{equation}
\Gamma ^{(l)}(p_{1},...,p_{\frac{l}{2}})=\mu ^{-2\epsilon }\underset{s=1}{%
\overset{\frac{l}{2}-1}{\dprod }}\dint \frac{d^{4}q_{s}}{(2\pi )^{4}}\frac{1%
}{\left( q_{s}^{2}-m_{0}^{2}\right) }\frac{1}{\left( (\overset{\frac{l}{2}}{%
\underset{i=1}{\dsum }}p_{i}-\overset{\frac{l}{2}-1}{\underset{i=1}{\dsum }}%
q_{i})^{2}-m_{0}^{2}\right) }  \label{apa6}
\end{equation}%
in last equation we can separate the integral in $d^{4}q_{\frac{l}{2}-1}$,
then%
\begin{equation}
\Gamma ^{(l)}(p_{1},...,p_{\frac{l}{2}})=\mu ^{-2\epsilon }\underset{s=1}{%
\overset{\frac{l}{2}-2}{\dprod }}\dint \frac{d^{4}q_{s}}{(2\pi )^{4}}\frac{1%
}{\left( q_{s}^{2}-m_{0}^{2}\right) }\dint\limits_{{}}^{{}}\frac{d^{4}q_{%
\frac{l}{2}-1}}{(2\pi )^{4}}\frac{1}{\left( q_{\frac{l}{2}%
-1}^{2}-m_{0}^{2}\right) }\frac{1}{\left( (r-q_{\frac{l}{2}%
-1})^{2}-m_{0}^{2}\right) }  \label{apa7}
\end{equation}%
where%
\begin{equation}
r=\overset{\frac{l}{2}}{\underset{i=1}{\dsum }}p_{i}-\overset{\frac{l}{2}-2}{%
\underset{i=1}{\dsum }}q_{i}  \label{apa8}
\end{equation}%
that is, we separate the coordinate $q_{\frac{l}{2}-1}$ on the sum in the
denominator of eq.(\ref{apa6}). The integral in $d^{4}q_{\frac{l}{2}-1}$ can
be computed where the result can be seen in \cite{Ramond}, page 122,
eq.(4.4.16).\footnote{%
There will be $l-1$ contributions corresponding to the different channels.}
For simplicity we will assume that the result can be written as%
\begin{equation}
\dint\limits_{{}}^{{}}\frac{d^{4}q_{\frac{l}{2}-1}}{(2\pi )^{4}}\frac{1}{%
\left( q_{\frac{l}{2}-1}^{2}-m_{0}^{2}\right) }\frac{1}{\left( (r-q_{\frac{l%
}{2}-1})^{2}-m_{0}^{2}\right) }=\underset{i=-1}{\overset{+\infty }{\sum }}%
\eta _{i}\epsilon ^{i}+\underset{i=0}{\overset{+\infty }{\sum }}%
R_{i}(p_{1},...,p_{\frac{l}{2}},q_{1},..,q_{\frac{l}{2}})\epsilon ^{i}
\label{apa9}
\end{equation}%
where $\eta _{i}$ are constants and $R_{i}$ are functions of the argument.
For example%
\begin{gather}
\eta _{-1}=-\frac{3}{16\pi ^{2}}\text{ \ \ \ \ }\eta _{0}=-\frac{3}{16\pi
^{2}}\left( \ln (\frac{4\pi }{m_{0}^{2}})-\gamma \right) \text{ \ \ \ \ \ }%
\eta _{1}=\frac{1}{768}+\frac{1}{32\pi ^{2}}\left[ \frac{\gamma }{2}-\ln (2%
\sqrt{\pi })\right] ^{2}  \label{apa10} \\
R_{0}(r)=\frac{1}{32\pi ^{2}}\sqrt{1+\frac{4m_{0}^{2}}{r^{2}}}\ln \left\{ 
\frac{\sqrt{1+\frac{4m_{0}^{2}}{r^{2}}}+1}{\sqrt{1+\frac{4m_{0}^{2}}{r^{2}}}%
-1}\right\}  \notag
\end{gather}%
Introducing the result of eq.(\ref{apa9}) in eq.(\ref{apa7}) we have%
\begin{equation}
\Gamma ^{(l)}(p_{1},...,p_{\frac{l}{2}})=\mu ^{-2\epsilon }\left[ \underset{%
i=-1}{\overset{+\infty }{\sum }}\eta _{i}\epsilon ^{i}\underset{s=1}{\overset%
{\frac{l}{2}-2}{\dprod }}\dint \frac{d^{4}q_{s}}{(2\pi )^{4}}\frac{1}{\left(
q_{s}^{2}-m_{0}^{2}\right) }+\underset{i=0}{\overset{+\infty }{\sum }}%
\epsilon ^{i}\underset{s=1}{\overset{\frac{l}{2}-2}{\dprod }}\dint \frac{%
d^{4}q_{s}}{(2\pi )^{4}}\frac{R_{i}}{\left( q_{s}^{2}-m_{0}^{2}\right) }%
\right]  \label{apa11}
\end{equation}%
But%
\begin{equation}
\underset{s=1}{\overset{\frac{l}{2}-2}{\dprod }}\dint \frac{d^{4}q_{s}}{%
(2\pi )^{4}}\frac{1}{\left( q_{s}^{2}-m_{0}^{2}\right) }=\dint%
\limits_{{}}^{{}}\frac{d^{4}q_{1}}{(2\pi )^{4}}\frac{1}{\left(
q_{1}^{2}-m_{0}^{2}\right) }\times ...\times \dint\limits_{{}}^{{}}\frac{%
d^{4}q_{\frac{l}{2}-2}}{(2\pi )^{4}}\frac{1}{\left( q_{\frac{l}{2}%
-2}^{2}-m_{0}^{2}\right) }=\left[ \Delta (0)\right] ^{\frac{l}{2}-2}
\label{apa12}
\end{equation}%
Then, if the second term in the r.h.s. of eq.(\ref{apa11}) do not contribute
with more poles in $\epsilon $ we can write 
\begin{equation}
\Gamma ^{(l)}(p_{1},...,p_{\frac{l}{2}})=\mu ^{-2\epsilon }\left[ \left[
\Delta (0)\right] ^{\frac{l}{2}-2}\underset{i=-1}{\overset{+\infty }{\sum }}%
\eta _{i}\epsilon ^{i}+\underset{i=0}{\overset{+\infty }{\sum }}\epsilon
^{i}f_{i}^{(l)}\right]  \label{apa13}
\end{equation}%
where\footnote{%
In eq.(\ref{apa14}), when $l=4$, the product start at $s=1$ and ends at $s=0$%
, in this case, it only remains the integrand, which reads $R_{i}$.}%
\begin{equation}
f_{i}^{(l)}=\underset{s=1}{\overset{\frac{l}{2}-2}{\dprod }}\dint \frac{%
d^{4}q_{s}}{(2\pi )^{4}}\frac{R_{i}}{\left( q_{s}^{2}-m_{0}^{2}\right) }
\label{apa14}
\end{equation}%
The following step is to use eq.(\ref{new5}) to compute the first term of
the r.h.s. of eq.(\ref{apa13}). Using eq.(\ref{new5}) we have%
\begin{equation}
\left[ \Delta (0)\right] ^{\frac{l}{2}-2}=\frac{1}{\epsilon ^{\frac{l}{2}-2}}%
\underset{j=0}{\overset{+\infty }{\sum }}\xi _{j-(\frac{l}{2}%
-2)}^{(l-2)}\epsilon ^{j}  \label{apa15}
\end{equation}%
Then%
\begin{equation}
\left[ \Delta (0)\right] ^{\frac{l}{2}-2}\underset{i=-1}{\overset{+\infty }{%
\sum }}\eta _{i}\epsilon ^{i}=\frac{1}{\epsilon ^{\frac{l}{2}-2}}\underset{%
n=0}{\overset{+\infty }{\sum }}\epsilon ^{n-1}S_{n}^{(l)}  \label{apa16}
\end{equation}%
where we use $\underset{n=0}{\sum }a_{n}\underset{n=0}{\sum }b_{n}=\underset{%
n=0}{\overset{+\infty }{\sum }}\overset{n}{\underset{k=0}{\sum }}%
a_{k}b_{n-k} $ and 
\begin{equation}
S_{n}^{(l)}=\underset{k=0}{\overset{n}{\sum }}\xi _{k-(\frac{l}{2}%
-2)}^{(l-2)}\eta _{n-k-1}  \label{apa17}
\end{equation}%
Replacing eq.(\ref{apa16})\ in eq.(\ref{apa13})\ we have%
\begin{equation}
\Gamma ^{(l)}(p_{1},...,p_{\frac{l}{2}})=\mu ^{-2\epsilon }\left[ \frac{1}{%
\epsilon ^{\frac{l}{2}-2}}\underset{n=0}{\overset{+\infty }{\sum }}\epsilon
^{n-1}S_{n}^{(l)}++\underset{i=0}{\overset{+\infty }{\sum }}\epsilon
^{i}f_{i}^{(l)}\right]  \label{apa18}
\end{equation}%
Finally we have to multiply the factor $\mu ^{-2\epsilon }$, to do it, we
expand it in Taylor series around $\epsilon =0$:%
\begin{equation}
\mu ^{-2\epsilon }=\underset{i=0}{\overset{+\infty }{\sum }}\frac{(-2\ln \mu
)^{j}}{j!}\epsilon ^{j}  \label{apa19}
\end{equation}%
then%
\begin{equation}
\Gamma ^{(l)}(p_{1},...,p_{\frac{l}{2}})=\frac{1}{\epsilon ^{\frac{l}{2}-2}}%
\underset{n=0}{\overset{+\infty }{\sum }}\epsilon ^{n-1}\underset{k=0}{%
\overset{n}{\sum }}\frac{(-2\ln \mu )^{k}}{k!}S_{n-k}^{(l)}+\underset{n=0}{%
\overset{+\infty }{\sum }}\epsilon ^{n}\underset{k=0}{\overset{n}{\sum }}%
\frac{(-2\ln \mu )^{k}}{k!}f_{n-k}^{(l)}  \label{apa20}
\end{equation}

The observable-state model projects over the finite contribution, which
corresponds to the term in eq.(\ref{apa20})\ with $\epsilon ^{0}$, then the
first term of the r.h.s. of last equation implies that $n-1=\frac{l}{2}-2$
and the second term implies that $n=0$, then%
\begin{equation}
\Gamma _{\text{finite}}^{(l)}(p_{1},...,p_{\frac{l}{2}})=\underset{k=0}{%
\overset{\frac{l}{2}-1}{\sum }}\frac{(-2\ln \mu )^{k}}{k!}S_{\frac{l}{2}%
-1-k}^{(l)}+f_{0}^{(l)}  \label{apa21}
\end{equation}

For example, with $l=4$ we obtain%
\begin{equation}
\Gamma _{\text{finite}}^{(4)}(p_{1},p_{2})=\underset{k=0}{\overset{1}{\sum }}%
\frac{(-2\ln \mu )^{k}}{k!}S_{1-k}^{(4)}+f_{0}^{(4)}=S_{1}^{(4)}-2\ln \mu
S_{0}^{(4)}+f_{0}^{(4)}  \label{apa22}
\end{equation}

Using eq.(\ref{apa14}) and eq.(\ref{apa17}), each term in the last equation
reads%
\begin{gather}
S_{0}^{(4)}=\underset{k=0}{\overset{0}{\sum }}\xi _{k}^{(2)}\eta _{-k-1}=\xi
_{0}^{(2)}\eta _{-1}=\eta _{-1}=-\frac{3}{16\pi ^{2}}  \label{apa23} \\
S_{1}^{(4)}=\underset{k=0}{\overset{1}{\sum }}\xi _{k}^{(2)}\eta _{-k}=\xi
_{0}^{(2)}\eta _{0}+\xi _{1}^{(2)}\eta _{-1}=\eta _{0}=-\frac{3}{16\pi ^{2}}%
\left( \ln (\frac{4\pi }{m_{0}^{2}})-\gamma \right)  \notag \\
f_{0}^{(4)}=R_{0}=\frac{1}{32\pi ^{2}}\sqrt{1+\frac{4m_{0}^{2}}{r^{2}}}\ln
\left\{ \frac{\sqrt{1+\frac{4m_{0}^{2}}{r^{2}}}+1}{\sqrt{1+\frac{4m_{0}^{2}}{%
r^{2}}}-1}\right\}  \notag
\end{gather}%
where we have used that $\xi _{0}^{(2)}=1$ and $\xi _{1}^{(2)}=0$ and eq.(%
\ref{apa10}).\footnote{%
The coefficients $\xi _{i}^{(2)}$ can be computed from eq.(\ref{new5}), in
the case $l=2$, $\left[ \Delta (0)\right] ^{0}=1$, so the only term finite
and different from zero is $\xi _{0}^{(2)}=1$.}

Then%
\begin{equation}
\Gamma _{\text{finite}}^{(4)}(p_{1},p_{2})=-\frac{3}{16\pi ^{2}}\left(
-\gamma +\ln (\frac{4\pi \mu ^{2}}{m_{0}^{2}})\right) +\frac{1}{16\pi ^{2}}%
\sqrt{1+\frac{4m_{0}^{2}}{r^{2}}}\ln \left\{ \frac{\sqrt{1+\frac{4m_{0}^{2}}{%
r^{2}}}+1}{\sqrt{1+\frac{4m_{0}^{2}}{r^{2}}}-1}\right\}  \label{apa24}
\end{equation}%
which coincide with eq.(4.4.16) of \cite{Ramond}.

\end{document}